%% file: DDPaper.tex
\documentclass[prd,tightenlines,floatfix,showpacs,preprintnumbers,nofootinbib,eqsecnum,superscriptaddress,twocolumn]{revtex4-1}

\usepackage{color}      
\usepackage{amsbsy}     
\usepackage{amsfonts}       
\usepackage{amsmath}        
\usepackage{amssymb}        
\usepackage{xcolor}
\usepackage{wasysym}
\usepackage{multirow}
\usepackage{mciteplus}
\usepackage{psfrag}
\usepackage{slashed}
\usepackage{cancel}
\usepackage{graphicx}
\usepackage{ulem}

\usepackage{hyperref}
 
\newcommand{\DMNLO}{{\tt DM@NLO}}
\newcommand{\MO}{{\tt micrOMEGAs}}
\newcommand{\DS}{{\tt DarkSUSY}}

\newcommand{\SPheno}{{\tt SPheno}}

\newcommand{\MSbar}{{$\overline{\mathrm{MS}}$}}
\newcommand{\DRbar}{{$\overline{\mathrm{DR}}$}}

\newcommand{\beq}{\begin{equation}}
\newcommand{\eeq}{\end{equation}}
\newcommand{\bea}{\begin{eqnarray}}
\newcommand{\eea}{\end{eqnarray}}
\def\Eins{{\mathchoice {\rm 1\mskip-4mu l} {\rm 1\mskip-4mu l}{\rm 1\mskip-4.5mu l} {\rm 1\mskip-5mu l}}}

\catcode`\@=11
\font\manfnt=manfnt
\def\Watchout{\@ifnextchar [{\W@tchout}{\W@tchout[1]}}
\def\W@tchout[#1]{{\manfnt\@tempcnta#1\relax%
  \@whilenum\@tempcnta>\z@\do{%
    \char"7F\hskip 0.3em\advance\@tempcnta\m@ne}}}
\let\foo\W@tchout
\def\dubious{\@ifnextchar[{\@dubious}{\@dubious[1]}}

\def\@dubious[#1]{%
  \color{red}\setbox\@tempboxa\hbox{\@W@tchout#1}
  \@tempdima\wd\@tempboxa
  \list{}{\leftmargin\@tempdima}\item[\hbox to 0pt{\hss\@W@tchout#1}]}
\def\@W@tchout#1{\W@tchout[#1]}
\catcode`\@=12

\begin{document}
\preprint{MS-TP-16-14}

\title{SUSY-QCD corrections for the direct detection of neutralino dark matter \\ and correlations with the relic density}

\author{M.~Klasen}
 \email{michael.klasen@uni-muenster.de}
 \affiliation{
	Institut f\"ur Theoretische Physik, Westf\"alische Wilhelms-Universit\"at M\"unster, Wilhelm-Klemm-Stra{\ss}e 9, D-48149 M\"unster, Germany
  }

\author{K.~Kova\v{r}\'ik}
 \email{karol.kovarik@uni-muenster.de}
 \affiliation{
	Institut f\"ur Theoretische Physik, Westf\"alische Wilhelms-Universit\"at M\"unster, Wilhelm-Klemm-Stra{\ss}e 9, D-48149 M\"unster, Germany
  }

\author{P.~Steppeler}
 \email{p\_step04@uni-muenster.de}
 \affiliation{
	Institut f\"ur Theoretische Physik, Westf\"alische Wilhelms-Universit\"at M\"unster, Wilhelm-Klemm-Stra{\ss}e 9, D-48149 M\"unster, Germany
  }

\date{\today}

\begin{abstract}
In this paper, we perform a full next-to-leading order (NLO) QCD calculation of neutralino scattering on protons or neutrons in the MSSM. We match the results of the NLO QCD calculation to the scalar and axial-vector operators in the effective field theory approach. These govern the spin-independent and spin-dependent detection rates, respectively. The calculations have been performed for general bino, wino and higgsino decompositions of neutralino dark matter and required a novel tensor reduction method of loop integrals with vanishing relative velocities and Gram determinants. Numerically, the NLO QCD effects are shown to be of at least of similar size and sometimes larger than the currently estimated nuclear uncertainties. We also demonstrate the interplay of the direct detection rate with the relic density when consistently analyzed with the program \DMNLO.
\end{abstract}


\maketitle

\input{intro.tex}

\input{analytical.tex}

\input{results.tex}

\input{conclusion.tex}

\acknowledgments

We thank A.~Crivellin, J.~Harz, B.~Herrmann, V.~Drach and S.~A.~Larin for fruitful discussions. This work was supported by Deutsche Forschungsgemeinschaft (DFG) under project KL 1266/5-1 and by the Helmholtz Alliance for Astroparticle Physics (HAP). All figures have been produced using {\tt Matplotlib} \cite{Hunter:2007}.

\appendix
\input{appendix.tex}

\input{bib.tex}

\end{document}

%% file: intro.tex
\section{Introduction}
\label{Intro}

Nowadays, the existence of dark matter is well established by experimental observations on many different length scales. In particular, on cosmological length scales, measurements of the temperature anisotropies of the Cosmic Microwave Background (CMB) allow a very precise determination of the relic density of dark matter. The most recent value obtained by the Planck collaboration \cite{Planck}, including polarization data from the Wilkinson Microwave Anisotropy Probe \cite{WMAP9}, is
\begin{equation}
 \Omega_{\mathrm{CDM}}h^2 = 0.1199 \pm 0.0022,
 \label{Planck}
\end{equation}
where $h$ denotes the present Hubble expansion rate in units of 100 km s$^{-1}$ Mpc$^{-1}$. Even though the quantity of dark matter in the Universe is known very accurately, its nature remains concealed. The reason for this unfortunate situation is that so far all experimental evidence for dark matter stems exclusively from its gravitational interaction.

Among the numerous attempts to explain dark matter, postulating the existence of a yet unknown Weakly Interacting Massive Particle (WIMP) is a widely adopted paradigm. This approach is attractive because a WIMP with typical weak scale interactions and a mass of $\sim$ 100 GeV naturally leads to the observed relic density via thermal freeze-out  \cite{Klasen:2015uma}. The canonical example for a WIMP is the lightest neutralino $\tilde{\chi}^0_1$, which is the lightest supersymmetric particle in many scenarios of the Minimal Supersymmetric Standard Model (MSSM). In the following, we refer to it simply by ``the neutralino''. Remember that giving rise to a suitable dark matter candidate is only a positive byproduct of introducing Supersymmetry (SUSY) as the most general space-time symmetry, which is furthermore motivated by its elegant solution to the hierarchy problem and the possible unification of gauge and Yukawa couplings. Alternatively, more minimal extensions to the SM with additional Higgs doublets \cite{Barbieri:2006dq}, neutrinos \cite{Ma:2006km} or other scalars and fermions \cite{Esch:2013rta} may be considered.

Assuming that dark matter actually consists of WIMPs, additional non-gravitational detection possibilities open up. First, one can try to directly produce WIMPs at a collider. As the WIMPs themselves are not detectable with current collider detectors, the typical observable of such a process consists of a single jet or gauge boson and missing transverse energy. The second possibility is to look for Standard Model annihilation products of WIMPs in very dense astronomical objects such as the Sun or the center of the Galaxy, where the WIMPs might have accumulated. The observational challenge of this indirect detection approach is to distinguish between the astrophysical background and a possible WIMP signal. Finally, one can try to observe the rare interactions of a WIMP with a nucleus by detecting its recoil in the so-called direct detection experiments. The technical difficulty here is to detect a very weak signal, while simultaneously excluding all non-dark matter sources \cite{Klasen:2015uma}. 

The direct detection rate, i.e.\ the number of events per time and per detector mass, depends on the dark matter-nucleus interaction. On the microscopic level, this corresponds to the interaction of the WIMP with the quarks and gluons inside the nucleons of the nucleus. However, as the typical process energies are much smaller than the mediator masses\footnote{As this condition is not necessarily fulfilled at a collider, EFT methods are under debate in this context, and so-called simplified models should be used \cite{Buchmueller, Busoni, DeSimone, Matsumoto}.} of the microscopic theory, it is customary to calculate the corresponding cross sections in the framework of effective field theories (EFT) \cite{Hill1, Hill2, HisanoEFT, JijiEFT}. In the EFT approach, the heavy particles which mediate the interaction between dark matter and the constituents of the nucleus are integrated out.

Integrating-out heavy particles translates different Lorentz structures of the microscopic theory to different effective contact interactions expressed in terms of effective operators. Not all of the effective interactions contribute in the non-relativistic limit which is relevant for direct detection. In the MSSM, the dominant effective operators for neutralino dark matter are the scalar operator $m_q\bar{\chi}\chi\bar{q}q$ and the axial-vector operator $\bar{\chi}\gamma_\mu\gamma_5\chi\bar{q}\gamma^\mu\gamma_5q$, as the vector and tensor operators vanish in the case of a Majorana fermion. These operators lead to coherent spin-independent (SI) and spin-dependent (SD) contributions, respectively.

The tree-level contributions of neutralino dark matter to these operators have first been calculated in Ref.\ \cite{Griest}. Since then, several improvements have been made by either including additional operators like e.g.\ gluon operators \cite{Drees, HisanoGluon} or by calculating electroweak radiative corrections for pure wino, higgsino or bino dark matter \cite{Hisanoelw, Berlin1, Berlin2}. 

In this paper, we perform a full $\mathcal{O}(\alpha_s)$ calculation for the two dominant operators listed above. In contrast to previous approaches, we allow for a general neutralino admixture and calculate the radiative corrections using fully general loop integrals. By doing so, we implement a second, loop-improved dark matter observable in our numerical package \DMNLO, the first one being the relic density \cite{DMNLOPage, AFunnel, ChiChi2qq1, ChiChi2qq2, ChiChi2qq3, NeuQ2qx1, NeuQ2qx2, QQ2xx, Scalepaper}. Combining these calculations allows to effectively constrain the MSSM parameter space and precisely predict the direct detection rate.

The remainder of this paper is organized as follows: In Sec.\ \ref{Technical}, we briefly remind the reader how the direct detection rate is computed in general, and we describe our renormalization scheme. We present the underlying EFT calculation, specify the matching of full and effective theory, and describe the running of the operators and their associated Wilson coefficients. In order to use the same tensor reduction method for our direct detection and relic density calculation, the tensor reduction method had to be modified to account for vanishing Gram determinants. As this technical aspect might be interesting on its own, we illustrate it separately in App.\ \ref{GramDet}. Our numerical results are then given in Sec.\ \ref{Numerics}. We analyze the impact of the radiative corrections and contrast them with the nuclear uncertainties. We also study the influence of the neutralino composition on the resulting neutralino-nucleus cross sections. Furthermore, we combine our direct detection and relic density routines to obtain precise predictions for the neutralino-nucleon cross section in a given scenario. Finally, we conclude in Sec.\ \ref{Conclusion}. We do not present any technical details of our relic density calculations here, but instead refer the reader to our previous papers and in particular Ref.\ \cite{ChiChi2qq3}.


%% file: analytical.tex
\section{Calculation of the neutralino nucleon cross section}
\label{Technical}

\subsection{Composition of direct detection rate}
\label{DDFormulas}

In this subsection, we briefly review the standard formulas for the calculation of neutralino direct detection rates. The desired quantity is the rate of events $\mathrm{d}R$ per energy interval $\mathrm{d}E$. This differential event rate is typically expressed in terms of counts per kg and day and keV. It can be written as
\beq
\label{DDRate2}
\frac{\mathrm{d}R}{\mathrm{d}E} = \sum_i c_i \frac{\sigma_i}{2m_{\tilde{\chi}^0_1}\mu_i^2}\rho_0 \eta_i.
\eeq
The sum runs over all detector nuclides $i$, and the factor $c_i$ denotes the mass fraction of the nuclear species $i$ in the detector. Let $m_i$ be the mass of the nucleus of species $i$. Then $\mu_i$ is the reduced mass
\beq
\mu_i = \frac{m_{\tilde{\chi}^0_1}m_i}{m_{\tilde{\chi}^0_1} + m_i}.
\eeq

The local dark matter density is described by $\rho_0$. Before using the canonical value of 0.3 GeV/cm$^3$, one should calculate the neutralino relic density to ensure that its value is in agreement with the experimental constraints and that the neutralinos can solely account for dark matter. $\eta_i$ contains the integration over the dark matter velocity relative to the detector $\vec{v}$,
\beq
\eta_i = \int_{v_{\mathrm{min},i}}^{v_{\mathrm{esc}}}\mathrm{d}^3v\frac{f(\vec{v})}{v}\quad\mathrm{with}\quad v_{\mathrm{min},i} = \sqrt{\frac{m_iE}{2\mu_i^2}}.
\eeq
The lower integration limit  $v_{\mathrm{min},i}$ is given by the minimal neutralino velocity, which can cause a recoil energy $E$. The upper integration limit is fixed by the galactical escape speed $v_{\mathrm{esc}}$, which is usually set to 544 km/s. Faster particles are not gravitationally bound in the Milky Way. More details on the integration limits can be found in Refs.\ \cite{SmithRAVE, vesc, CirelliTools}. $f(\vec{v})$ is the local velocity distribution, which is typically assumed to be Maxwellian. However, several studies have unveiled that this simplification might not describe the situation properly, see e.g. \cite{vDistrib, GreenDistribution, DD_ProtonNeutron}. All the particle physics is contained in the cross sections for elastic nucleus-neutralino scattering $\sigma_i$, where we distinguish between spin-independent and spin-dependent contributions. 

The spin-independent cross section can be written as
\beq
\label{sigmaSI}
\sigma_i^{\mathrm{SI}} = \frac{\mu_i^2}{\pi}\left|Z_i g_p^{\mathrm{SI}} +(A_i-Z_i)g_n^{\mathrm{SI}}\right|^2|F_i^{\mathrm{SI}}(Q_i)|^2,
\eeq
where $F_i^{\mathrm{SI}}(Q_i)$ is the spin-independent structure function for the nucleus $i$. It depends on the momentum transfer $Q_i = \sqrt{2m_iE}$, can be understood as the Fourier transform of the nucleon density, and is normalised to $F_i^{\mathrm{SI}}(0) = 1$. The nucleus $i$ consists of $Z_i$ protons and $A_i -Z_i$ neutrons, where $Z_i$ is its atomic number and $A_i$ is its mass number. To enable a comparison of direct detection results, that is independent of the detector material and technology, the experimental collaborations typically publish constraints on the cross section of the dark matter particle and a single nucleon\footnote{At this point, the typical assumption is that the interaction strength of neutralinos is the same for protons and neutrons. This is not necessarily fulfilled in a non-minimal model like the MSSM. Therefore we keep our calculations general and distinguish between protons and neutrons.} $N$, which simply reads
\beq
\label{sigmaSINucleon}
\sigma_N^{\mathrm{SI}} = \frac{\mu_N^2}{\pi}\left|g_N^{\mathrm{SI}}\right|^2.
\eeq
Here, the neutralino-nucleus reduced mass $\mu_i$ is replaced by the neutralino-nucleon reduced mass $\mu_N$ in complete analogy. The nucleon masses $m_N$ are given by
\beq
\label{Nucleonmasses}
m_p = 0.9383\ \mathrm{GeV}\quad\mathrm{and}\quad m_n = 0.9396\ \mathrm{GeV}.
\eeq

The effective spin-independent four-fermion couplings among neutralinos and protons $p$ or neutrons $n$ are denoted by $g_p^{\mathrm{SI}}$ and $g_n^{\mathrm{SI}}$. They can be determined via
\beq
g_N^{\mathrm{SI}} = \sum_q \langle N |\bar{q}q| N\rangle \alpha_{q}^{\mathrm{SI}},
\eeq
where the nucleon index $N$ stands either for a proton or a neutron and where the sum runs over all quark types $q$.\footnote{We are summing over all quark types, as we do not include gluon operators yet. Alternatively, one could replace the heavy-quark contributions by loop-induced gluon processes including heavy quarks as virtual particles.} The spin-independent interaction between quarks and neutralinos is denoted by $\alpha_{q}^{\mathrm{SI}}$. The quark matrix element $\langle N |\bar{q}q| N\rangle$ can be qualitatively understood as the probability to find a quark $q$ in the nucleon $N$. We write it as
\beq
\label{fTDef}
\langle N |m_q \bar{q}q| N\rangle = f_{Tq}^N m_N,
\eeq
where $m_N$ denotes the nucleon mass and $m_q$ the quark mass. The scalar coefficients $f_{Tq}^N$ are determined experimentally or via lattice QCD. We point out that especially $f_{Ts}^N$ is affected by experimental uncertainties, which mainly stem from the determination of the pion-nucleon sigma term \cite{NuclearUncertainties, EllisHadron, BottinoNucleon}. We use the values  given in Refs.\ \cite{CrivellinNucleon, Hoferichter, Junnarkar} which differ from the ones implemented in \DS\ \cite{DarkSUSY} or \MO\ \cite{micrOMEGAs}. We list all values for comparison in Tab.\ \ref{fTTable}.\footnote{We are working with \MO\ \texttt{2.4.1} to benefit from our established relic density interface. However, we have updated the nuclear input values to the most recent version manually. Hence the values given in Tab.\ \ref{fTTable} correspond to \MO\ \texttt{4.2.5}.} The factors $f_{Tq}^N$ of the heavy quarks are linked to those of the light quarks via \cite{Shifman}
\beq
f_{Tc}^N = f_{Tb}^N = f_{Tt}^N= \frac{2}{27}\left(1-\sum_{q=u,d,s} f_{Tq}^N\right).
\eeq

\begin{table}
\caption{Scalar coefficients $f_{Tq}^N$ used in different codes.}
\begin{center}
\begin{tabular}{|c|ccc|}
\hline
Scalar coefficient & \DMNLO & \DS & \MO\\ 
\hline 
$f_{Tu}^p$  & 0.0208 & 0.023 &  0.0153 \\
$f_{Tu}^n$ & 0.0189 & 0.019 & 0.0110 \\
$f_{Td}^p$  & 0.0411 & 0.034 & 0.0191\\
$f_{Td}^n$ & 0.0451 & 0.041 & 0.0273\\
$f_{Ts}^p = f_{Ts}^n$  & 0.043 & 0.14 & 0.0447\\
$f_{Tc}^p = f_{Tb}^p = f_{Tt}^p$ & 0.0663 & 0.0595 & 0.0682\\
$f_{Tc}^n = f_{Tb}^n = f_{Tt}^n$ & 0.0661 & 0.0592 & 0.0679\\
\hline
\end{tabular} 
\end{center}
\label{fTTable}
\end{table}

The spin-dependent cross section can be cast into the form
\bea
\label{sigmaSD}
\sigma_i^{\mathrm{SD}} & = & \frac{4\mu_i^2}{2J +1}\big(|g_p^{\mathrm{SD}}|^2S_{\mathrm{pp},i}(Q_i) + |g_n^{\mathrm{SD}}|^2S_{\mathrm{nn},i}(Q_i)\nonumber\\
&&  + |g_p^{\mathrm{SD}}g_n^{\mathrm{SD}}|S_{\mathrm{pn},i}(Q_i)\big),
\eea
where $J$ denotes the nuclear spin. Details on the spin structure functions $S_{\mathrm{pp},i}(Q_i)$, $S_{\mathrm{nn},i}(Q_i)$ and $S_{\mathrm{pn},i}(Q_i)$ can be found in Ref.\ \cite{VogelStructure}. The spin-dependent cross section for a neutralino and a single nucleon $N$ reads
\beq
\label{sigmaSDNucleon}
\sigma_N^{\mathrm{SD}} = \frac{3\mu_N^2}{\pi}|g_N^{\mathrm{SD}}|^2.
\eeq

The effective spin-dependent four-fermion couplings among neutralinos and protons $p$ ($g_p^{\mathrm{SD}}$) or neutrons $n$ ($g_n^{\mathrm{SD}}$) are given by
\beq
g_N^{\mathrm{SD}} = \sum_{q= u,d,s}  (\Delta q)_N \alpha_{q}^{\mathrm{SD}}.
\eeq
In contrast to the spin-independent case, we sum only over the light quarks $u$, $d$ and $s$, as mainly these flavors contribute to the spin of the nucleon.\footnote{Note, however, that it was recently claimed that bottom quarks may also contribute to the spin-dependent interaction \cite{LiBottom}.} $(\Delta q)_N$ can be seen as the fraction of the nucleon spin carried by the quark $q$. More precisely, it describes the second moment of the polarized quark density and is related to the nucleon spin vector $s_\mu$ via
\beq
\langle N | \bar{q}\gamma_\mu\gamma_5 q| N\rangle = 2s_\mu(\Delta q)_N.
\eeq
We choose the default values of \MO\ for the polarized quark densitites
\bea
(\Delta u)_p = (\Delta d)_n & = & 0.842,\label{Delta1}\\
(\Delta d)_p = (\Delta u)_n & = & -0.427,\label{Delta2}\\
(\Delta s)_p = (\Delta s)_n & = & -0.085,
\eea
constrained by isospin symmetry, i.e.\ $(\Delta u)_p = (\Delta d)_n$ and $(\Delta d)_p = (\Delta u)_n$.

\subsection{Renormalization scheme}
\label{Renormalization}

Our QCD calculations at next-to-leading order (NLO) and beyond are performed within a hybrid on-shell/$\overline{\rm DR}$ renormalization scheme, described in detail in Refs.\ \cite{ChiChi2qq3, NeuQ2qx1, NeuQ2qx2}. In the quark sector, the top and bottom quark masses are defined on-shell and in the $\overline{\rm DR}$ scheme, respectively. Note that through the Yukawa coupling to (in particular the neutral pseudoscalar) Higgs boson resonances, the bottom quark mass can have a sizeable influence on the dark matter annihilation cross section and must therefore be treated with particular care. We obtain it from the SM $\overline{\rm MS}$ mass $m_b(m_b)$, determined in an analysis of $\Upsilon$ sum rules, through evolution to the scale $\mu_R$, transformation to the SM $\overline{\rm DR}$ and then MSSM $\overline{\rm DR}$ scheme \cite{ChiChi2qq3, NeuQ2qx1}. In the squark sector, we have five independent parameters 
\begin{equation}
 m_{\tilde{t}_1}, \quad m_{\tilde{b}_1}, \quad m_{\tilde{b}_2}, \quad A_t \quad\mathrm{and}\quad A_b=0.
 \label{eq:RenInput}
\end{equation}
The lighter stop mass and the two sbottom masses are taken to be on-shell, while the stop and  sbottom trilinear coupling parameters are taken in the $\overline{\rm DR}$ scheme. From these parameters, we compute as dependent quantities the stop and sbottom mixing angles $\theta_{\tilde{t}}$ and $\theta_{\tilde{b}}$ and $m_{\tilde{t}_2}$ for the heavier stop \cite{NeuQ2qx1}. The masses of the first- and second-generation squarks are taken on-shell. The strong coupling constant $\alpha_s(\mu_R)$ is renormalized in the MSSM $\overline{\rm DR}$ scheme with six active flavors and obtained after evolution of the world-average, five-flavor SM $\overline{\rm MS}$ value at the $Z^0$-boson mass to the renormalization scale $\mu_R$ and an intermediate transformation to the SM $\overline{\rm DR}$ scheme \cite{NeuQ2qx2}.


Although EFT calculations are usually performed in a minimal scheme such as \MSbar\ or its SUSY equivalent \DRbar, we continue to use the hybrid scheme presented above for three main reasons: First, we want to combine our direct detection calculations with our relic density analysis, where this scheme has proven very reliable. In particular, the on-shell description of the top quark leads to improved perturbative stability and better fits our supersymmetric processes and top quark final states in comparison to a definition in the \DRbar\ scheme \cite{Scalepaper}. A second reason is that the hybrid scheme also leads to improved perturbative stability for direct detection as described below in Sec.\ \ref{ScenarioC}. The last reason is that using this hybrid scheme allows for simpler comparison of the leading order result with \MO. This is due to fact that both, our calculation and \MO, use the same on-shell squark masses calculated by \SPheno\ as described in section \ref{Numerics}.

\subsection{Matching of the full and effective theory}

\begin{figure}
	\includegraphics[width=0.49\textwidth]{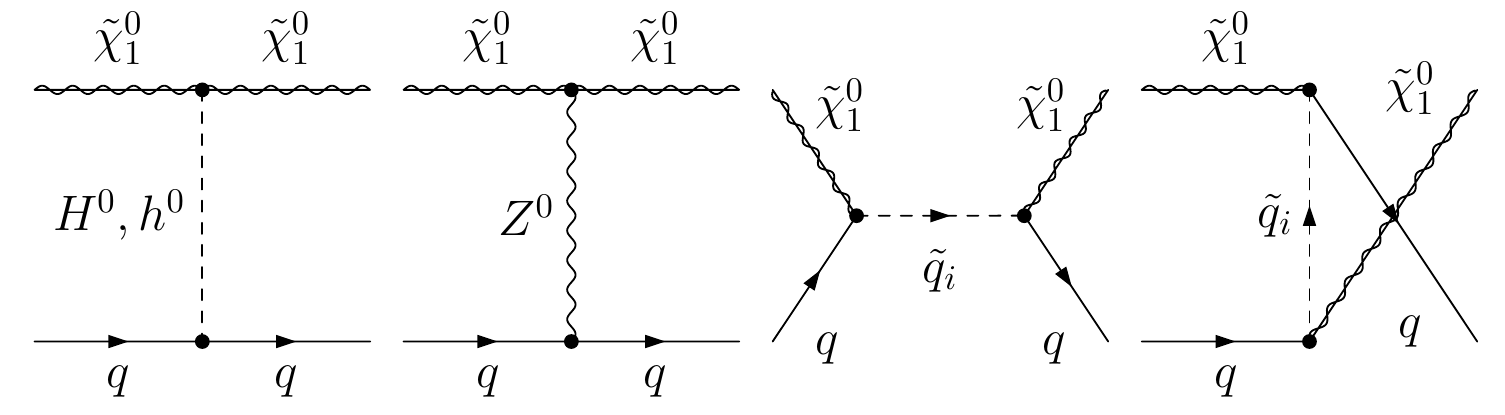}
	\caption{Tree-level processes in the full theory.}
	\label{fig:DDTree}
\end{figure}

\begin{figure}
	\includegraphics[width=0.18\textwidth]{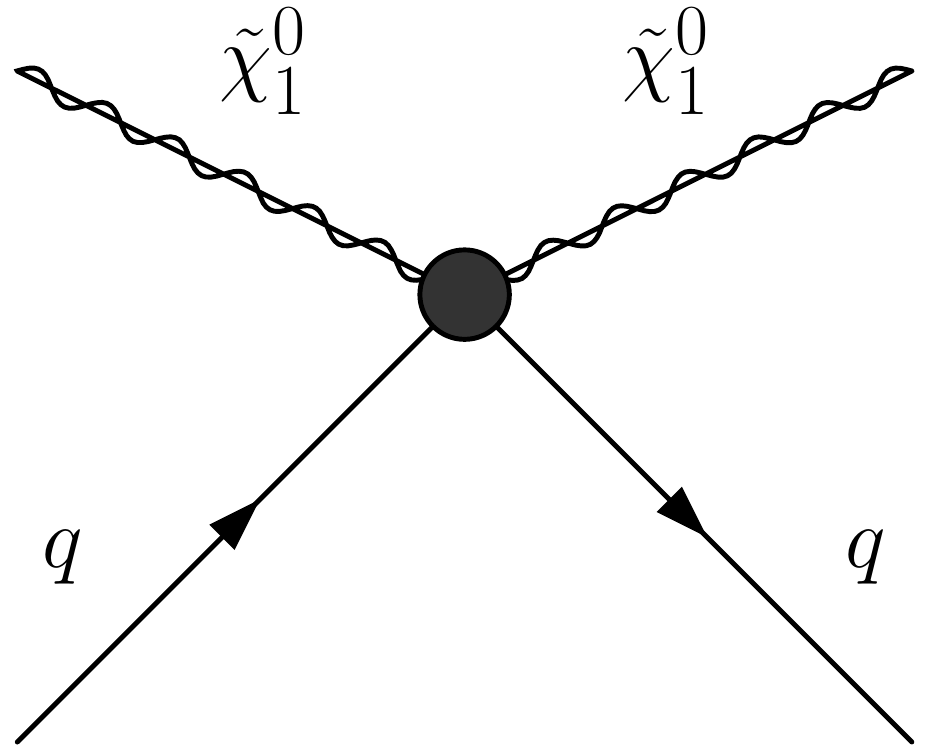}
	\includegraphics[width=0.18\textwidth]{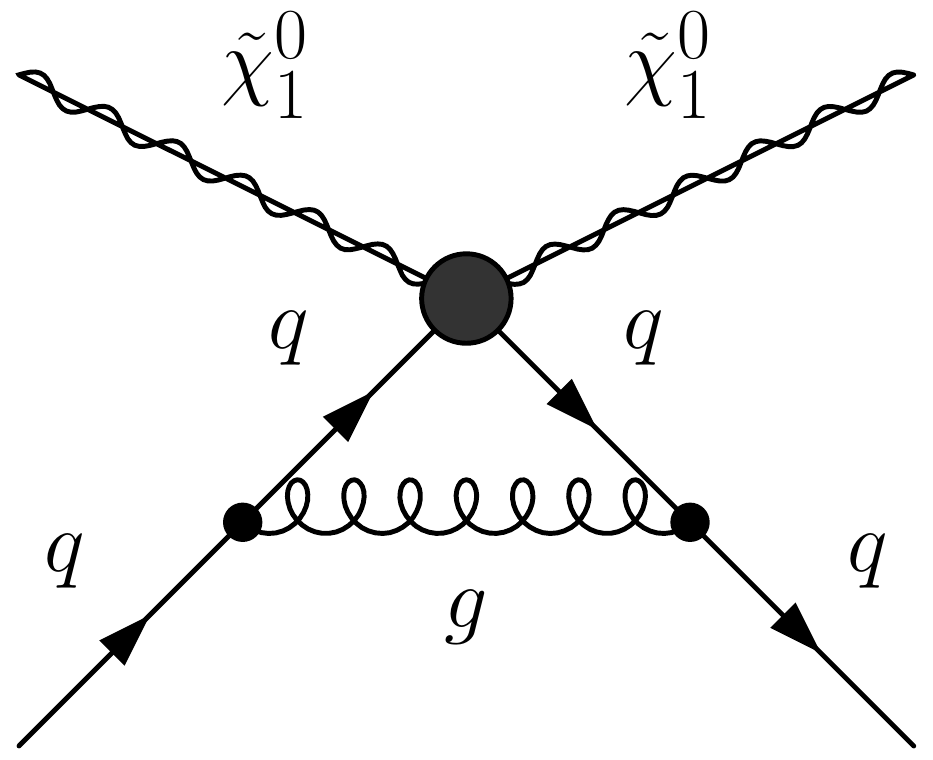}
	\caption{Tree-level process (left) and virtual correction (right) in the effective theory.}
	\label{fig:DDEFT}
\end{figure}	

This subsection is devoted to the matching of the full theory, namely the MSSM, valid at high energies ($\mu_\mathrm{high}\sim 1$ TeV) onto the effective energy valid at low energies ($\mu_\mathrm{low}\sim 5$ GeV). The tree-level diagrams of the scattering process $\tilde{\chi}^0_1q\rightarrow\tilde{\chi}^0_1q$ within the MSSM  are shown in Fig.\ \ref{fig:DDTree}. The corresponding amplitudes have to be evaluated at vanishing relative velocity and mapped onto the yet unknown Wilson coefficients $c_1$ and $c_2$ of the effective Lagrangian
\beq
\label{Leff}
\mathcal{L}_\mathrm{eff} = c_1Q_1 + c_2Q_2 = c_1\bar{\chi}\chi\bar{q}q + c_2\bar{\chi}\gamma_\mu\gamma_5\chi\bar{q}\gamma^\mu\gamma_5q.
\eeq
We stress that in this convention a factor $m_q$ has to be factored out of $c_1$ when replacing the nuclear matrix elements via Eq.\ (\ref{fTDef}). Both of the operators $Q_1$ and $Q_2$ given above lead to an effective four-fermion interaction as shown in the left diagram of Fig.\ \ref{fig:DDEFT}. The Higgs processes contribute solely to the scalar operator and the $Z^0$ processes solely to the axial-vector operator. We include only the scalar Higgs bosons $h^0$ and $H^0$ and not the pseudoscalar Higgs boson $A^0$, since the latter leads to the kinematically suppressed operator $\bar{\chi}\gamma_5\chi\bar{q}\gamma_5q$. The squark processes contribute to both operators. To bring the spinor fields into the desired order, a Fierz transformation has to be performed in this case. 

The aforementioned mapping onto the Wilson coefficients is governed by the matching condition. This condition demands that the amplitude of the full theory $\mathcal{M}_\mathrm{full}$ is reproduced by the effective theory at the high scale $\mu_\mathrm{high}$. At tree level we have
\beq
\mathcal{M}_\mathrm{full}^\mathrm{tree} \overset{!}{=} \mathcal{M}_\mathrm{eff}^\mathrm{tree} = c_1^\mathrm{tree}Q_1^\mathrm{tree} + c_2^\mathrm{tree}Q_2^\mathrm{tree}\label{TreeMatching}
\eeq
which leads to
\bea
c_1^\mathrm{tree} = \alpha_q^{\mathrm{SI}} & = & \sum_{\phi = h^0,H^0}\frac{g_{\tilde{\chi}\tilde{\chi}\phi}^Rg^L_{qq\phi}}{m_\phi^2} - \frac{1}{4}\sum_{i=1}^2\frac{g_{\tilde{\chi}\tilde{q}_iq}^Lg_{\tilde{\chi}\tilde{q}_iq}^{R*}}{m_{\tilde{q}_i}^2 - s}\nonumber\\
&& - \frac{1}{4}\sum_{i=1}^2\frac{g_{\tilde{\chi}\tilde{q}_iq}^Lg_{\tilde{\chi}\tilde{q}_iq}^{R*}}{m_{\tilde{q}_i}^2 - u},\\
c_2^\mathrm{tree} = \alpha_q^{\mathrm{SD}} & = & \frac{1}{2}\frac{g_{\tilde{\chi}\tilde{\chi}Z^0}^R(g^L_{qqZ^0} - g^R_{qqZ^0})}{m_{Z^0}^2}\nonumber\label{CSITree}\\
&& + \frac{1}{8}\sum_{i=1}^2\frac{|g_{\tilde{\chi}\tilde{q}_iq}^L|^2 + |g_{\tilde{\chi}\tilde{q}_iq}^{R}|^2}{m_{\tilde{q}_i}^2 - s}\nonumber\\
&& + \frac{1}{8}\sum_{i=1}^2\frac{|g_{\tilde{\chi}\tilde{q}_iq}^L|^2 + |g_{\tilde{\chi}\tilde{q}_iq}^{R}|^2}{m_{\tilde{q}_i}^2 - u}.\label{CSDTree}
\eea
In the limit of vanishing relative velocity, the Mandelstam variables $s$ and $u$ simplify to $(m_{\tilde{\chi}^0_1} \pm m_q)^2$, respectively. The elementary couplings between three particles $a$, $b$ and $c$ are denoted by $g_{abc}$. Using the chirality projectors $P_{L/R} = (\Eins \mp \gamma_5)/2$, they can be decomposed into left- and right-handed parts via
\beq
g_{abc} = g_{abc}^LP_L + g_{abc}^RP_R.
\eeq
Explicit expressions for the couplings can be found, e.g., in Ref.\ \cite{DreesBook}. The tree-level results have been analytically compared with those implemented in \DS. Taking into account that \DS\ does not distinguish between $s$- and $u$-channels, we find perfect agreement.

\begin{figure}
	\includegraphics[width=0.49\textwidth]{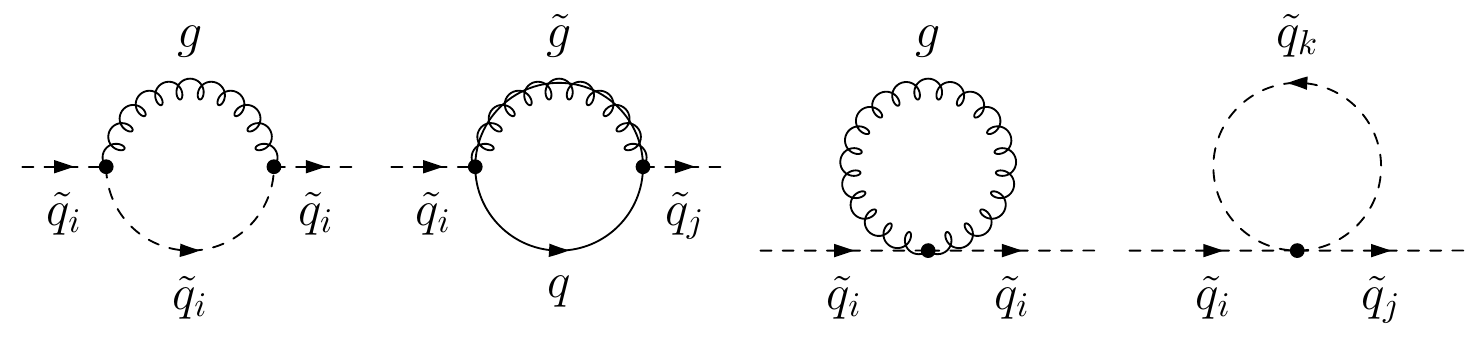}
	\includegraphics[width=0.49\textwidth]{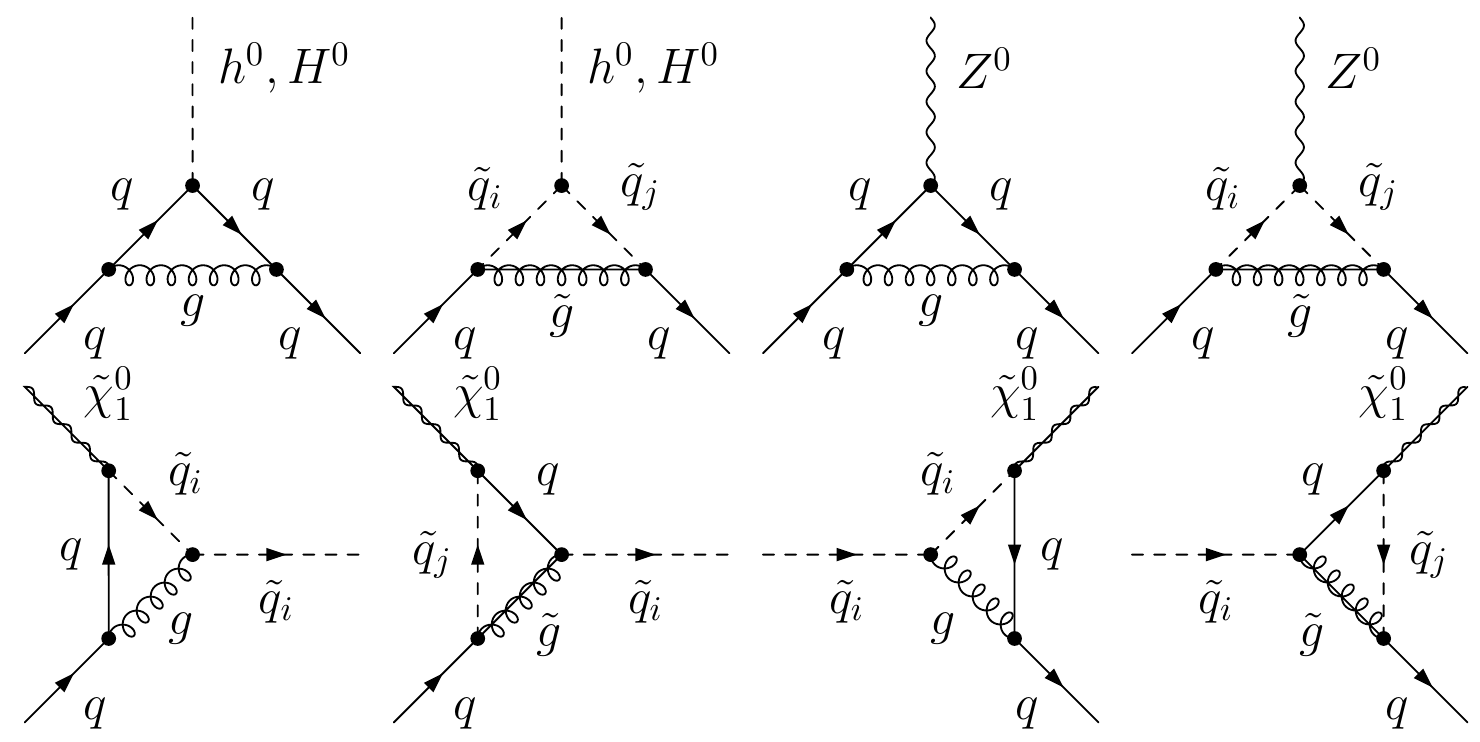}
	\includegraphics[width=0.49\textwidth]{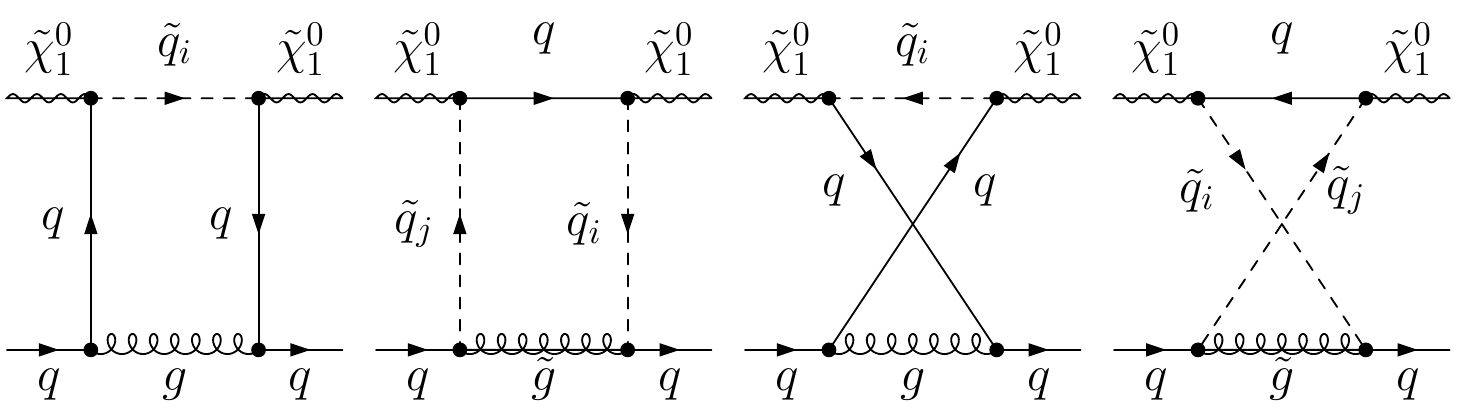}
	\caption{Virtual corrections in the full theory.}
	\label{fig:DDCorrections}
\end{figure}	

So far we have basically reproduced already available results. The next step is to improve on the tree-level calculation by including all $\mathcal{O}(\alpha_s)$ corrections to the leading operators. The corresponding diagrams within the full theory are shown in Fig.\ \ref{fig:DDCorrections}. We distinguish between propagator corrections (the first row), vertex corrections (the second and the third row) and box contributions (the last row).

We have calculated all of the loop amplitudes in full generality using dimensional reduction. The gluon propagator correction shown as the third diagram in the first row then vanishes, as it is proportional to the scaleless scalar integral $A_0(0) = 0$. In the case of the other propagator and the vertex corrections, we were able to benefit from previous loop calculations performed in the context of Ref.\ \cite{ChiChi2qq3}. The box amplitudes were calculated from scratch. These amplitudes lead to a plethora of effective operators. We keep only the most relevant, namely those of Eq.\ (\ref{Leff}). In case of the gluon boxes, Fierz transformations are necessary again. Explicit expressions for all involved loop amplitudes will be given in Ref.\ \cite{myPhD}.
In contrast to our relic density calculations, the loops are evaluated at zero relative velocity in the context of direct detection. This leads to additional problems, namely vanishing Gram determinants. We illustrate this technical issue separately in App.\ \ref{GramDet}. 

The propagator and vertex corrections give rise to ultraviolet divergences. These divergences are removed via renormalization (cf.\ Sec.\ \ref{Renormalization}), i.e.\ by adding the corresponding counterterms. A detailed description of the counterterms involved here is given in Ref.\ \cite{NeuQ2qx1}. As we always distinguish ultraviolet and infrared poles ($\epsilon_{UV}$ and $\epsilon_{IR}$) when evaluating loop integrals, we were able to explicitly check the ultraviolet safety of our calculation. 

Having the renormalized amplitudes of the full theory at hand, we can start with the matching procedure at NLO. The matching condition remains basically unchanged and reads
\bea\hspace*{-1cm}
\mathcal{M}_\mathrm{full}^\mathrm{NLO} & \overset{!}{=} & \mathcal{M}_\mathrm{eff}^\mathrm{NLO}.\\
\hspace*{-1cm}\Leftrightarrow\mathcal{M}_\mathrm{full}^\mathrm{tree} + \mathcal{M}_\mathrm{full}^\mathrm{1loop} & \overset{!}{=} & c_1^\mathrm{NLO}Q_1^\mathrm{NLO} + c_2^\mathrm{NLO}Q_2^\mathrm{NLO}.\label{NLOMatching}
\eea
In this convention, the full NLO result consists of the tree-level result and its $\mathcal{O}(\alpha_s)$ one-loop correction. The latter includes all the virtual corrections depicted in Fig.\ \ref{fig:DDCorrections}. The crucial point in Eq. (\ref{NLOMatching}) is that there is a one-loop correction to the Wilson coefficients \textit{and} the effective operators. We neglect terms of $\mathcal{O}(\alpha_s^2)$ and write
\begin{widetext}
\bea
\mathcal{M}_\mathrm{full}^\mathrm{tree} + \mathcal{M}_\mathrm{full}^\mathrm{1loop} & \overset{!}{=} & (c_1^\mathrm{tree} + c_1^\mathrm{1loop})(Q_1^\mathrm{tree}+ Q_1^\mathrm{1loop}) + (c_2^\mathrm{tree} + c_2^\mathrm{1loop})(Q_2^\mathrm{tree} + Q_2^\mathrm{1loop})\nonumber\\
& = & c_1^\mathrm{tree}Q_1^\mathrm{tree} + c_2^\mathrm{tree}Q_2^\mathrm{tree} + c_1^\mathrm{1loop}Q_1^\mathrm{tree} + c_2^\mathrm{1loop}Q_2^\mathrm{tree} + c_1^\mathrm{tree}Q_1^\mathrm{1loop} + c_2^\mathrm{tree}Q_2^\mathrm{1loop}.
\eea
\end{widetext}
At $\mathcal{O}(\alpha_s^0)$, we reproduce the tree-level matching condition Eq. (\ref{TreeMatching}). At $\mathcal{O}(\alpha_s)$, we obtain
\bea
&&\mathcal{M}_\mathrm{full}^\mathrm{1loop} - c_1^\mathrm{tree}Q_1^\mathrm{1loop} - c_2^\mathrm{tree}Q_2^\mathrm{1loop}\nonumber\\
&=& c_1^\mathrm{1loop}Q_1^\mathrm{tree} + c_2^\mathrm{1loop}Q_2^\mathrm{tree}.
\eea
Before we can calculate the $\mathcal{O}(\alpha_s)$ corrections to the Wilson coefficients, i.e.\ determine $c_1^\mathrm{1loop}$ and $c_2^\mathrm{1loop}$, we have to identify the one-loop corrections to the effective operators $Q_1^\mathrm{1loop}$ and $Q_2^\mathrm{1loop}$. These can be written as
\bea
Q_1^\mathrm{1loop} & = & (\mathcal{K}_\mathrm{EFTV1} + \mathcal{K}_\mathrm{EFTVC1})Q_1^\mathrm{tree}\quad\mathrm{and}\quad\\
Q_2^\mathrm{1loop} & = & (\mathcal{K}_\mathrm{EFTV2} + \mathcal{K}_\mathrm{EFTVC2})Q_2^\mathrm{tree},
\eea
i.e.\ they can be expressed as the tree-level operators multiplied with correction factors describing vertex corrections and vertex counterterms in the effective field theory. The vertex correction in the effective field theory is depicted on the right of Fig.\ \ref{fig:DDEFT}. This allows us to explicitly write down the one-loop Wilson coefficients as
\bea
c_1^\mathrm{1loop} & = & \alpha^\mathrm{SI}_{q,\mathrm{P}} + \alpha^\mathrm{SI}_{q,\mathrm{PC}} + \alpha^\mathrm{SI}_{q,\mathrm{V}} + \alpha^\mathrm{SI}_{q,\mathrm{VC}}\nonumber\\
& + &  \alpha^\mathrm{SI}_{q,\mathrm{B}} -  c_1^\mathrm{tree}(\mathcal{K}_\mathrm{EFTV1} + \mathcal{K}_\mathrm{EFTVC1}).\\
c_2^\mathrm{1loop} & = & \alpha^\mathrm{SD}_{q,\mathrm{P}} + \alpha^\mathrm{SD}_{q,\mathrm{PC}} + \alpha^\mathrm{SD}_{q,\mathrm{V}} + \alpha^\mathrm{SD}_{q,\mathrm{VC}}\nonumber\\
& + &  \alpha^\mathrm{SD}_{q,\mathrm{B}} -  c_2^\mathrm{tree}(\mathcal{K}_\mathrm{EFTV2} + \mathcal{K}_\mathrm{EFTVC2}).
\eea

Here $\alpha^\mathrm{SI}_{q,\mathrm{P}}$, $\alpha^\mathrm{SI}_{q,\mathrm{PC}}$, $\alpha^\mathrm{SI}_{q,\mathrm{V}}$, $\alpha^\mathrm{SI}_{q,\mathrm{VC}}$ and $\alpha^\mathrm{SI}_{q,\mathrm{B}}$ denote the contributions to the spin-independent four-fermion coupling stemming from the propagator corrections, propagator counterterm, vertex correction, vertex counterterms and box diagrams, respectively. The spin-dependent contributions are labeled analogous. All of these terms will be given explicitly in Ref. \cite{myPhD}.

We stress that $\alpha^\mathrm{SI}_{q,\mathrm{P}} + \alpha^\mathrm{SI}_{q,\mathrm{PC}}$, $\alpha^\mathrm{SI}_{q,\mathrm{V}} + \alpha^\mathrm{SI}_{q,\mathrm{VC}}$, $\alpha^\mathrm{SI}_{q,\mathrm{B}}$ and $\mathcal{K}_\mathrm{EFTV1} + \mathcal{K}_\mathrm{EFTVC1}$ are separately ultraviolet finite, and the same holds for the spin-dependent case and the associated correction factors $\mathcal{K}_\mathrm{EFTV2} + \mathcal{K}_\mathrm{EFTVC2}$. However, there are also infrared divergences involved, which have not been discussed yet. Although most of the individual terms given above are infrared divergent, $c_1^\mathrm{1loop}$ and $c_2^\mathrm{1loop}$ as a whole are infrared finite, which is an essential feature of the matching procedure. The appearance of infrared divergences is connected with massless particles like gluons. These particles are likewise degrees of freedom in the full and the effective theory. In other words: The infrared regime of both theories is the same. Whenever there occurs an infrared divergence in the full theory, the very same infrared divergence occurs in the effective theory as well, and both cancel during the matching procedure. In our calculation, this cancellation is due to the correction factors $\mathcal{K}_\mathrm{EFTV1}$, $\mathcal{K}_\mathrm{EFTVC1}$, $\mathcal{K}_\mathrm{EFTV2}$ and $\mathcal{K}_\mathrm{EFTVC2}$ which we list now.

The vertex correction factor $\mathcal{K}_\mathrm{EFTV1}$ is obtained by calculating the diagram shown on the right of Fig.\ \ref{fig:DDEFT} involving the effective operator $Q_1^\mathrm{tree}$. We get
\bea
\mathcal{K}_\mathrm{EFTV1} & = & \frac{\alpha_sC_F}{4\pi}\Big(4B_0 - 2  + 4p_bp_2(C_0 + C_1 + C_2)\Big),\nonumber\\
\eea
where the two- and three-point functions possess the arguments $B = B(p_b - p_2, m_q^2, m_q^2)$ and $C = C(p_2, p_b, 0, m_q^2, m_q^2)$. Here the four-momentum of the ingoing quark is denoted by $p_b$ and that of the outgoing quark by $p_2$. In the limit of vanishing relative velocity, we simply have $p = p_b = p_2$. Moreover, $C_F = 4/3$ denotes the usual color factor. This vertex correction is algebraically identical to the Higgs-gluon vertex shown on the very left in the second row of Fig.\ \ref{fig:DDEFT}, which has two important consequences. On the one hand, the Higgs-gluon vertex completely cancels in the matching procedure. The gluon is likewise a degree of freedom in the full and the effective theory and therefore the corresponding vertex correction occurs in both theories. It is included in the effective operator, not the Wilson coefficient. Moreover the correction factor $\mathcal{K}_\mathrm{EFTV1}$ is ultraviolet divergent, as it includes the two-point function $B_0$. To allow for a consistent matching procedure, we have to renormalize the effective theory in the same way as the full theory. This means that we have to add a counterterm $\delta c_1$ to the four-fermion coupling. This counterterm has to be of the same form as $\delta g_{\phi qq}$ (with $\phi = h^0,H^0$) and reads
\beq
\delta c_1^L = c_1^{\mathrm{tree},L}\left(\frac{\delta Z_m}{m_q} + \frac{1}{2}\delta Z_q^L + \frac{1}{2}\delta Z_q^{R*}\right),
\eeq
where $\delta Z_m$ denotes the mass and $\delta Z_q$ the wave function counterterm. For more details on these counterterms we refer the reader again to Ref.\ \cite{NeuQ2qx1}. The associated right-handed part of $\delta c_1$ is obtained by the substitution $L\leftrightarrow R$. The correction factor $\mathcal{K}_\mathrm{EFTVC1}$ is then simply given by
\beq
\mathcal{K}_\mathrm{EFTVC1} = \frac{\delta c_1^L/c_1^{\mathrm{tree},L} + \delta c_1^R/c_1^{\mathrm{tree},R}}{2}.
\eeq

Remember that $c_1^\mathrm{tree}$ does not only incorporate Higgs contributions, but that squark processes contribute as well (cf.\ Eq.\ (\ref{CSITree})). Whereas the Higgs-gluon vertex correction and its associated counterterm completely vanish in the matching procedure, this is not true for the vertex corrections to the squark processes shown in the third row of Fig.\ \ref{fig:DDCorrections} and their counterterms. However, the infrared divergences of these corrections and the ones stemming from the boxes shown in the last row of Fig.\ \ref{fig:DDCorrections} are precisely cancelled by the correction factors. This is an important consistency check of the whole calculation. Thanks to our generic implementation of loop integrals and the discrimination between ultraviolet and infrared poles, we could verify this cancellation explicitly.

We continue with the determination of $\mathcal{K}_\mathrm{EFTV2}$, i.e.\ the vertex correction factor for the spin-dependent operator $Q_2$. The associated diagram is shown on the right of Fig. \ref{fig:DDEFT} again, the only difference to the previous case is the included four-fermion coupling. Keeping only the relevant effective operators, we obtain
\bea
\mathcal{K}_\mathrm{EFTV2} & = & \frac{\alpha_sC_F}{4\pi}\Big(2B_0 + 4p_bp_2(C_0 + C_1 + C_2)\nonumber\\
&&  - 4C_{00} -1\Big)
\eea
where the two- and three-point functions possess the same arguments as before. The missing piece is the counterterm $\delta c_2$, which renders the vertex correction given above ultraviolet finite. This counterterm is constructed in analogy to $\delta g_{Z^0qq}$ and reads
\beq
\delta c_2^L = c_2^{\mathrm{tree},L}\left(\frac{1}{2}\delta Z_q^{\mathrm{SM},L} + \frac{1}{2}\delta Z_q^{\mathrm{SM},L*} + \frac{\alpha_sC_F}{\pi}\right).
\label{EFTVertexCSD}
\eeq
As before, the correction factor $\mathcal{K}_\mathrm{EFTVC2}$ is obtained via
\beq
\mathcal{K}_\mathrm{EFTVC2} = \frac{\delta c_2^L/c_2^{\mathrm{tree},L} + \delta c_2^R/c_2^{\mathrm{tree},R}}{2}.
\eeq
Note that we have included the additional finite part $\frac{\alpha_sC_F}{\pi}$ to retain a conventional axial current divergence which is in agreement with Refs. \cite{Hill2} and \cite{Larin}.\footnote{The results given in Ref. \cite{Larin} were obtained using the \MSbar\ scheme and dimensional regularization. Transferring results from this scheme to the \DRbar\ scheme and dimensional reduction -- which we are using -- is nontrivial in general. Discrepancies may arise due to the treatment of $\gamma_5$ in $D$ dimensions. However, these problems should occur at the three-loop order for the first time and do neither affect the finitite contribution included in Eq. (\ref{EFTVertexCSD}) nor the running of the axial-vector operator presented in the next section \cite{LarinMail}.} Moreover we incorporate just Standard Model contributions to $\delta Z_q$ in this case. The reason is as follows: In case of the Higgs vertex corrections including the gluon and the gluino, only the former is ultraviolet divergent. The whole counterterm $\delta g_{\phi qq}$ is responsible for the cancellation of this divergence. As the gluon vertex correction occurs likewise in the effective theory, we have constructed its associated counterterm $\delta c_1$ in complete analogy to $\delta g_{\phi qq}$. In case of the $Z^0$ vertex corrections including the gluon and the gluino, both are ultraviolet divergent. The divergences of the first diagram are removed by the Standard Model part of $\delta g_{Z^0 qq}$ and the latter by the SUSY part of $\delta g_{Z^0 qq}$. During the matching procedure, the gluon vertex correction and its corresponding counterterm has to cancel, whereas the vertex correction including the gluino and its counterterm contributes to the Wilson coefficient. Hence we only include Standard Model contributions to the spinor field counterterms in $\delta c_2$. This completes our matching calculation at NLO. 

\subsection{Running of effective operators and associated Wilson coefficients}
\label{RunningSection}

The matching calculation presented in the last subsection is performed at the high scale $\mu_\mathrm{high}\sim 1$ TeV. In contrast, the nuclear matrix elements are defined at a low scale $\mu_\mathrm{low}\sim 5$ GeV. This is the energy regime we finally aim to describe with our effective field theory. To connect the two energy regimes, we have to evolve the effective operators and associated Wilson coefficients from the high scale down to the low scale by solving the corresponding renormalization group equations (RGEs). This part of the calculation is briefly referred to as ``running'' and is presented in this subsection.

The scale dependence of the Wilson coefficients is inverse to that of the corresponding operators. Therefore it cancels in the product, which is an essential feature of any operator product expansion. In the effective Lagrangian introduced in Eq.\ (\ref{Leff}), we have neglected higher-dimensional operators in our operator product expansion, i.e.\
\beq
\mathcal{L}_\mathrm{eff} =  \sum_{i=1}^\infty c_iQ_i \approx c_1\bar{\chi}\chi\bar{q}q + c_2\bar{\chi}\gamma_\mu\gamma_5\chi\bar{q}\gamma^\mu\gamma_5q +\ldots
\eeq
As we are interested only in QCD effects, the running of the two operators given above is solely determined by their respective quark parts.

The scalar operator $m_q\bar{q}q$ is scale independent. As a consequence, the running calculation in the spin-indepependent case is rather simple. We have to factor out the quark mass $m_q(\mu_\mathrm{high})$ from the coefficient $c_1$. This quark mass has to be evolved down to the low scale $\mu_\mathrm{low}$ in the usual way, i.e.\ by solving its RGE. We then replace the combination $m_q(\mu_\mathrm{low})\bar{q}q$ via Eq.\ (\ref{fTDef}).

In contrast to that, the renormalization and the resulting running of the axial-vector operator is not trivial. This calculation has first been performed in Ref.\ \cite{Larin}. The relevant renormalization constant reads
\bea
Z_A^\mathrm{Singlet} & = & 1 + \frac{\alpha_s}{\pi}C_F - \frac{1}{\epsilon_{UV}}\left(\frac{\alpha_s}{4\pi}\right)^2\left(\frac{20}{9}n_f + \frac{88}{3}\right)\nonumber\\
&&  + \mathcal{O}(\alpha_s^3),
\eea
where $n_f$ denotes the number of active flavors and an additional finite term has been included to cure the axial anomaly. It is precisely this term which has been included in Eq.\ (\ref{EFTVertexCSD}) as well. Finite terms of order $\mathcal{O}(\alpha_s^2)$ have been neglected, as they are irrelevant for the running up to the desired order. Given this constant, we can calculate the corresponding anomalous dimension via
\beq
\gamma_A^\mathrm{Singlet} = (Z_A^\mathrm{Singlet})^{-1}\frac{\mathrm{d}}{\mathrm{d}\log\mu}Z_A^\mathrm{Singlet}
\eeq
and obtain
\beq
\gamma_A^\mathrm{Singlet} = \left(\frac{\alpha_s}{4\pi}\right)^2 16n_f + \mathcal{O}(\alpha_s^3).
\eeq
To arrive at this result one has to insert the RGE of the strong coupling constant including its divergent part, namely
\beq
\frac{\mathrm{d}g}{\mathrm{d}\log\mu} = -\epsilon_{UV}g + \beta(g),
\eeq
where $\beta(g)$ is the usual QCD beta function
\bea
\frac{\beta(g)}{g} & = & -\beta_0\frac{\alpha_s}{4\pi} + \mathcal{O}(\alpha_s^2) = -(11-\frac{2}{3}n_f)\frac{\alpha_s}{4\pi} + \mathcal{O}(\alpha_s^2).\nonumber\\
\eea
The remaining step is to determine the running of the Wilson coefficient $c_2$ via
\beq
\frac{\mathrm{d}}{\mathrm{d}\log\mu}c_2(\mu) = \gamma_A^\mathrm{Singlet}c_2(\mu).
\eeq
We finally obtain
\beq
\frac{c_2(\mu_\mathrm{low})}{c_2(\mu_\mathrm{high})} = \exp\left(\frac{2n_f(\alpha_s(\mu_\mathrm{high}) - \alpha_s(\mu_\mathrm{low}))}{\beta_0\pi}\right),
\eeq
which agrees with the result given in Ref.\ \cite{Hill2}. Note that in general different operators may mix under renormalization. This is fortunately not the case here, but it will happen when one includes e.g.\ the gluon operator $G_{\mu\nu}G^{\mu\nu}$ \cite{Hill2}.

%% file: results.tex
\section{Numerical results}
\label{Numerics}

\begin{table*}
	\caption{pMSSM input parameters for three selected reference scenarios. All parameters except $\tan\beta$ are given in GeV. }
	\begin{tabular}{|c|ccccccccccc|}
		\hline
			$\quad$ & $\quad\tan\beta\quad$ & $\quad\mu\quad$ & $\quad m_A\quad$ & $\quad M_1\quad$ & $\quad M_2\quad$ & $\quad M_3\quad$ & 		$\quad M_{\tilde{q}_{1,2}}\quad$ & $\quad M_{\tilde{q}_3}\quad$ & $\quad M_{\tilde{u}_3}\quad$ & $\quad M_{\tilde{\ell}}\quad$& $\quad A_t\quad$ \\ 
			\hline 
			A & 13.4 & 1286.3 & 1592.9 & 731.0 & 766.0 & 1906.3 & 3252.6 & 1634.3 & 1054.4 & 3589.6 & -2792.3\\			
			B & 13.7 & 493.0 & 500.8 & 270.0 & 1123.4 & 1020.3 & 479.9 & 1535.5 & 836.7 & 3469.4 & -2070.9\\	
			C &  7.0 & 815.0 & 1452.8 & 675.3 & 1423.4 & 1020.3 & 809.9 & 1835.5 & 1436.7 & 3469.4 & -2670.9\\			
			\hline
	\end{tabular}
	\label{ScenarioList}
\end{table*}
	
\begin{table*}
	\caption{Gaugino and squark masses and other selected observables corresponding to the reference scenarios of Tab.\ \ref{ScenarioList}. All masses are given in GeV.}
	\begin{tabular}{|c|cc|cc|ccccc|ccc|}
		\hline
			$\quad$  & ~~ $m_{\tilde{\chi}^0_1}$~~ & ~~$m_{\tilde{\chi}^0_2}$~~ &  ~~$m_{\tilde{\chi}^{\pm}_1}$~~ & ~~$m_{\tilde{\chi}^{\pm}_2}$~~ & ~~$m_{\tilde{u}_1}$~~ & ~~$m_{\tilde{d}_1}$~~ & ~~$m_{\tilde{t}_1}$~~ & ~~$m_{\tilde{b}_1}$~~ & ~~$m_{\tilde{g}}$~~ & ~~$m_{h^0}$~~ & ~~$\Omega_{\tilde{\chi}^0_1} h^2$~~ & $\mathrm{BR}(b\rightarrow s\gamma)$ \\
			\hline 
			A & 738.1 & 802.4 & 802.3 & 1295.1 & 3270.9 & 3271.6 & 993.9 & 1622.9 & 2049.9 & 126.3 & 0.1244 & $3.0\cdot 10^{-4}$ \\
			B & 265.7 & 498.4 & 495.7 & 1135.3 & 549.5 & 555.7 & 802.9 & 1531.0 & 1061.2 & 124.8 & 0.1199 & $3.6\cdot 10^{-4}$\\
			C & 669.2 & 826.6 & 819.6 & 1438.9 & 865.0 & 868.4 & 1389.1 & 1832.3 & 1090.7 & 125.2 & 0.1179 & $3.3\cdot 10^{-4}$\\	
			\hline		
	\end{tabular}
	\label{ScenarioProps}
\end{table*}

\begin{table}
	\caption{Most relevant (co)annihilation channels in the reference scenarios of Tab.\ \ref{ScenarioList}. Channels which contribute less than 1\% to the thermally averaged cross section and/or are not implemented in our code are not shown.}
	\begin{tabular}{|rl|cccc|}
		\hline
		 &  & ~~~~ A ~~~~ & ~~~~ B ~~~~ & ~~~~ C ~~~~ &  \\
		\hline
		$\tilde{\chi}^0_1 \tilde{\chi}^0_1 \to$ & $t\bar{t}$ & 1\% & 10\% & 52\% & \\
		                                        & $b\bar{b}$ & 9\% & 78\% &  40\% & \\

		$\tilde{\chi}^0_1 \tilde{\chi}^0_2 \to$ & $t\bar{t}$ &  3\% & &  & \\
		                                        & $b\bar{b}$ & 23\% &   &  & \\

		$\tilde{\chi}^0_1 \tilde{\chi}^{\pm}_1 \to$ & $t\bar{b}$ & 43\% & & & \\
		\hline
		\multicolumn{2}{|c|}{Total} & 79\% & 88\% & 92\% & \\
		\hline
	\end{tabular}
	\label{ScenarioChannels}
\end{table}

In this section we describe our numerical setup and present numerical results for three selected reference scenarios. These scenarios are defined in a phenomenological MSSM (pMSSM) with eleven free parameters, which we have already used in our previous analyses. This setup was designed for relic density calculations including light stops \cite{NeuQ2qx1, NeuQ2qx2, QQ2xx}. As it has proven sufficient for finding interesting direct detection scenarios, we stick to it for consistency and keep in mind, that a more specific pMSSM setup may lead to considerably larger loop contributions. 

The aforementioned eleven free parameters are as follows: The Higgs sector is fixed by the higgsino mass parameter $\mu$, the ratio of the vacuum expectation values of the two Higgs doublets $\tan\beta$, and the pole mass of the pseudoscalar Higgs boson $m_A$. The gaugino sector is defined by the bino ($M_1$), wino ($M_2$) and gluino ($M_3$) mass parameters, which in our setup are not related through any assumptions stemming from Grand Unified Theories. Moreover we define a common soft SUSY-breaking mass parameter $M_{\tilde{q}_{1,2}}$ for the first- and second-generation squarks. The third-generation squark masses are controlled by the parameter $M_{\tilde{q}_3}$ associated with sbottoms and left-handed stops and by the parameter $M_{\tilde{u}_3}$ for right-handed stops. The trilinear coupling in the stop sector is given by $A_t$, while the trilinear couplings of the other sectors, including $A_b$, are set to zero. Since the slepton sector is not at the center of our attention, it is parametrized by a single soft parameter $M_{\tilde{\ell}}$. The most interesting parameters for the following discussion are those determining the neutralino decomposition ($\mu$, $M_1$ and $M_2$) and $M_{\tilde{q}_{1,2}}$. 


These eleven pMSSM input parameters are defined in the \DRbar\ scheme at the scale $\tilde{M} = 1$ TeV according to the SPA convention \cite{Spa}. We identify this scale with our renormalization scale $\mu_{R}$, which simultaneously corresponds to the high scale $\mu_{\mathrm{high}}$ of our EFT calculation. The input parameters are handed over to the numerical package \SPheno\ \cite{SPheno} to calculate the associated physical spectrum.

We neglect the masses of the quarks of the first two generations in the kinematics to improve numerical stability. On the other hand, we keep those masses in the Yukawa couplings to allow for Higgs exchange processes. Remember that the Yukawa masses are basically factored out of the amplitudes and replaced by the nuclear matrix elements via Eq.\ (\ref{fTDef}). It has been checked explicitly that the effect of this simplification on the final results is negligible. 

Our three reference scenarios are listed in Tab.\ \ref{ScenarioList}. Table \ref{ScenarioProps} contains the corresponding relevant gaugino and squark masses\footnote{We are not showing the squark masses $m_{\tilde{u}_2}$, $m_{\tilde{d}_2}$, $m_{\tilde{c}_1}$, $m_{\tilde{c}_2}$, $m_{\tilde{s}_1}$ and $m_{\tilde{s}_2}$. However, as we are working with a common soft mass parameter $M_{\tilde{q}_{1,2}}$, all squark masses of the first two generations are roughly the same.} as well as the obtained mass of the lightest neutral (and thus SM-like) Higgs boson, the neutralino relic density computed at tree level with \MO\ and the important branching ratio of the rare $B$-meson decay $b\to s\gamma$ computed with \SPheno. Moreover, Tab.\ \ref{ScenarioChannels} lists the most relevant (co)annihilation channels for determining the relic density. Other important parameters are the neutralino mixing angles, i.e.\ its bino, wino and higgsino admixture. As the phenomenology of the three reference scenarios is to a large extent driven by these parameters, we explore them in more detail in the following. We devote an individual subsection to each scenario.

\subsection*{Scenario A -- Bino-wino dark matter}

We start by investigating scenario A. This scenario has been introduced in Ref.\ \cite{ChiChi2qq3} and studied again in Ref.\ \cite{Scalepaper}. Its main feature are sizeable gaugino coannihilation contributions to the relic density calculation as listed in Tab.\ \ref{ScenarioChannels}. The direct detection in this scenario is in no way special and we include this scenario as an arbitrary conservative case.

\begin{figure}
	\includegraphics[width=0.49\textwidth]{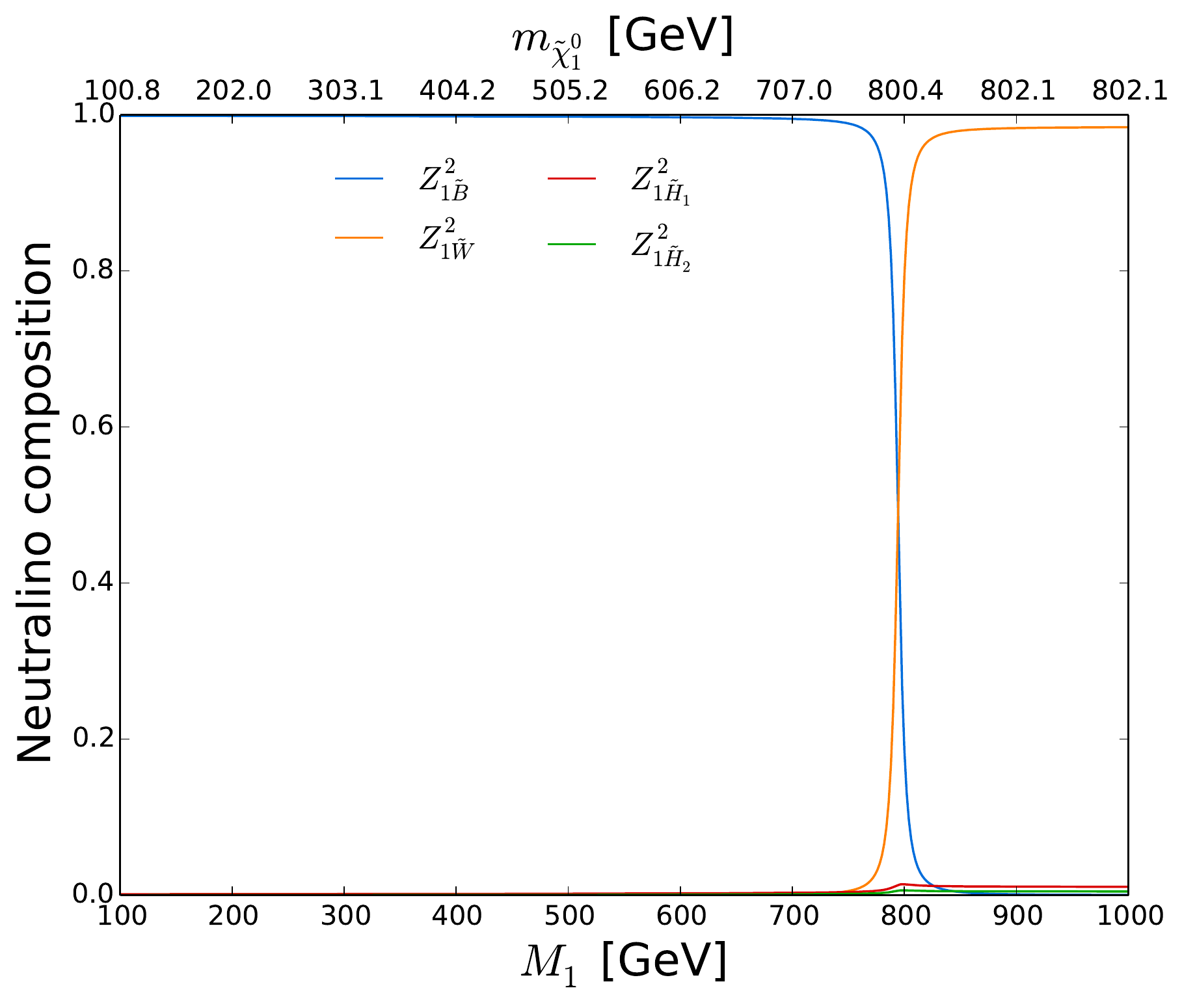}
	\caption{Neutralino decomposition in scenario A.}
	\label{fig:MixingScenA}
\end{figure}

The decomposition of the neutralino in dependence of the pMSSM input parameter $M_1$ is shown in Fig.\ \ref{fig:MixingScenA}. As long as $M_1<M_2$, the neutralino is mostly bino. It turns into mostly wino when $M_1 > M_2 = 766$ GeV while the higgsino content always stays small due to $M_1, M_2 < \mu$. Note that scenario A itself, i.e the cosmologically preferred region, sits near the turnover ($M_1 = $ 731 GeV). This situation is encountered in many pMSSM scenarios and clearly calls for a general treatment of the neutralino admixture. We also show the associated neutralino mass on the top of each plot as a derived parameter. This connects our theoretical predictions to experimental exclusion limits, which are usually given in dependence of the WIMP mass. Note that the correspondance between $M_1$ and $m_{\tilde{\chi}^0_1}$ is basically 1:1 for $M_1$ up to 800 GeV, but for larger values of $M_1$ the neutralino becomes mostly wino, so that its mass is almost independent of $M_1$. 

\begin{figure*}
	\includegraphics[width=0.49\textwidth]{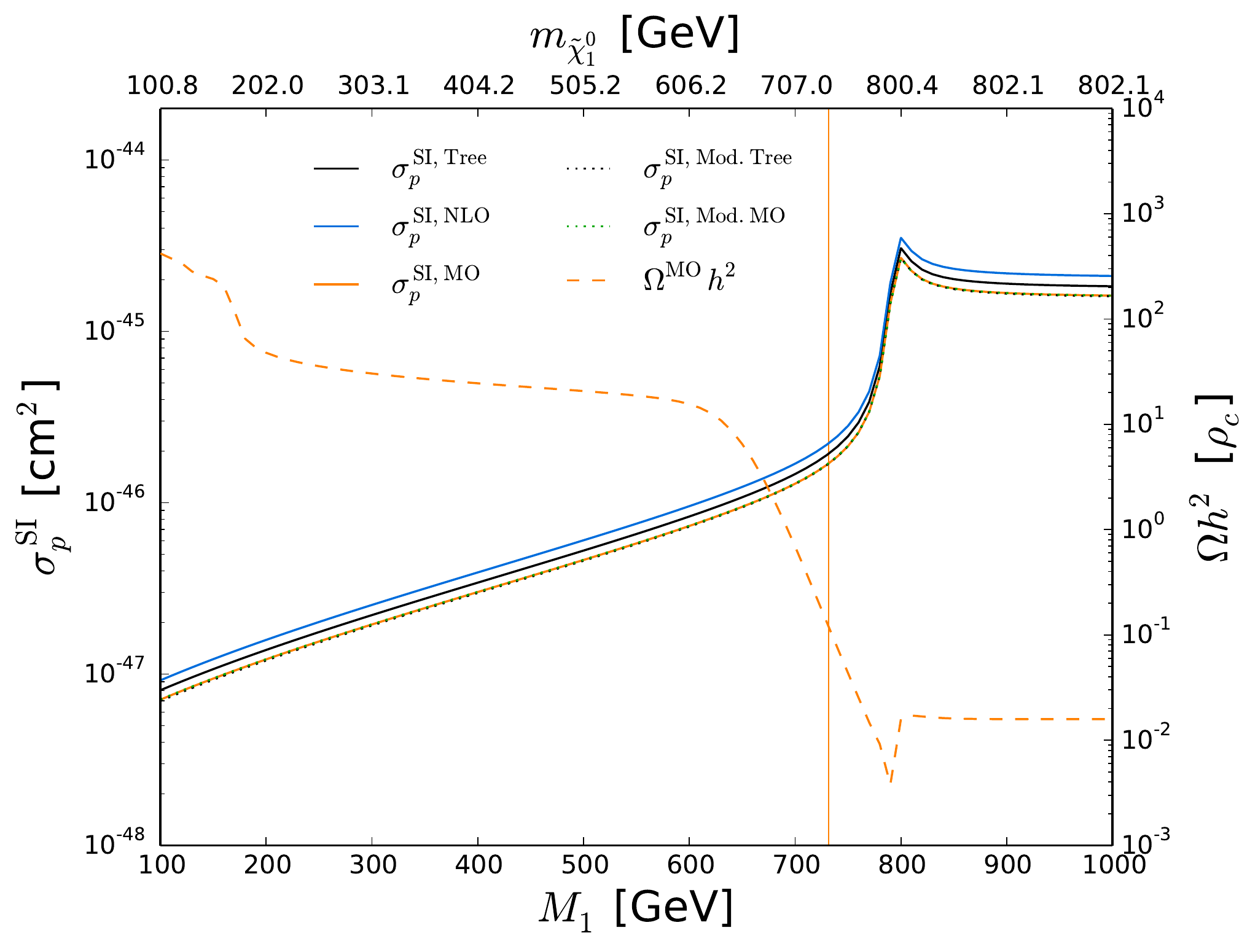}
	\includegraphics[width=0.49\textwidth]{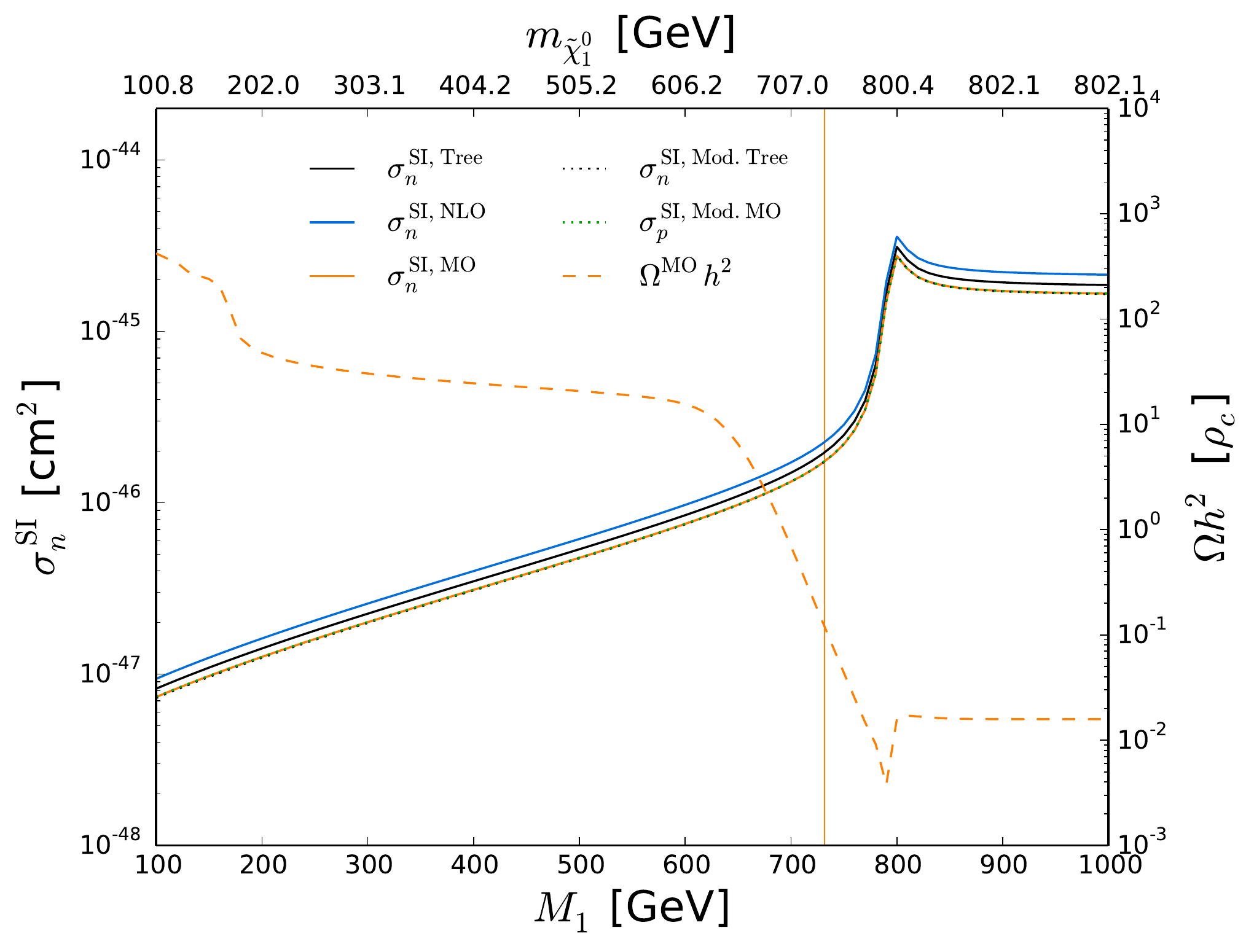}
	\includegraphics[width=0.49\textwidth]{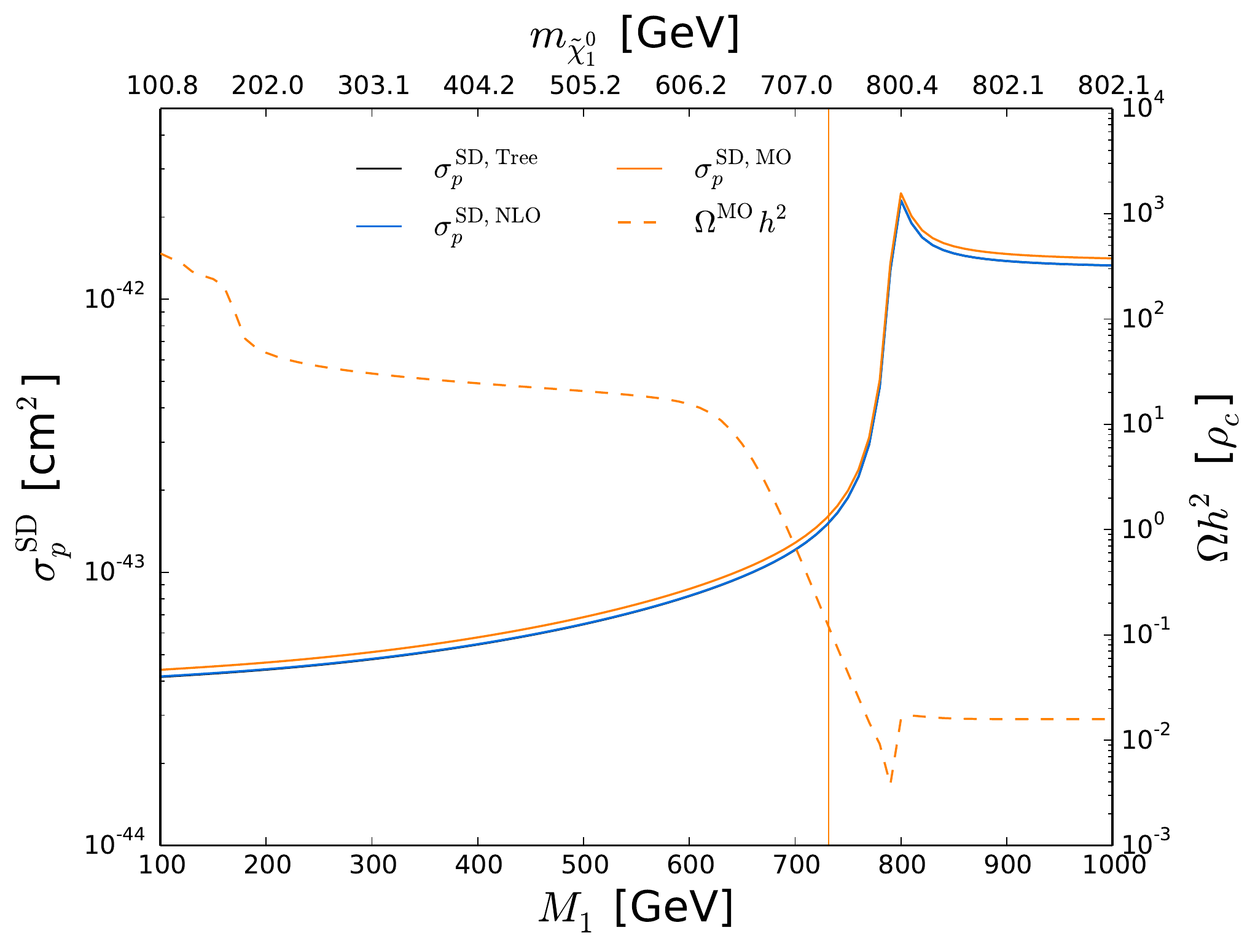}
	\includegraphics[width=0.49\textwidth]{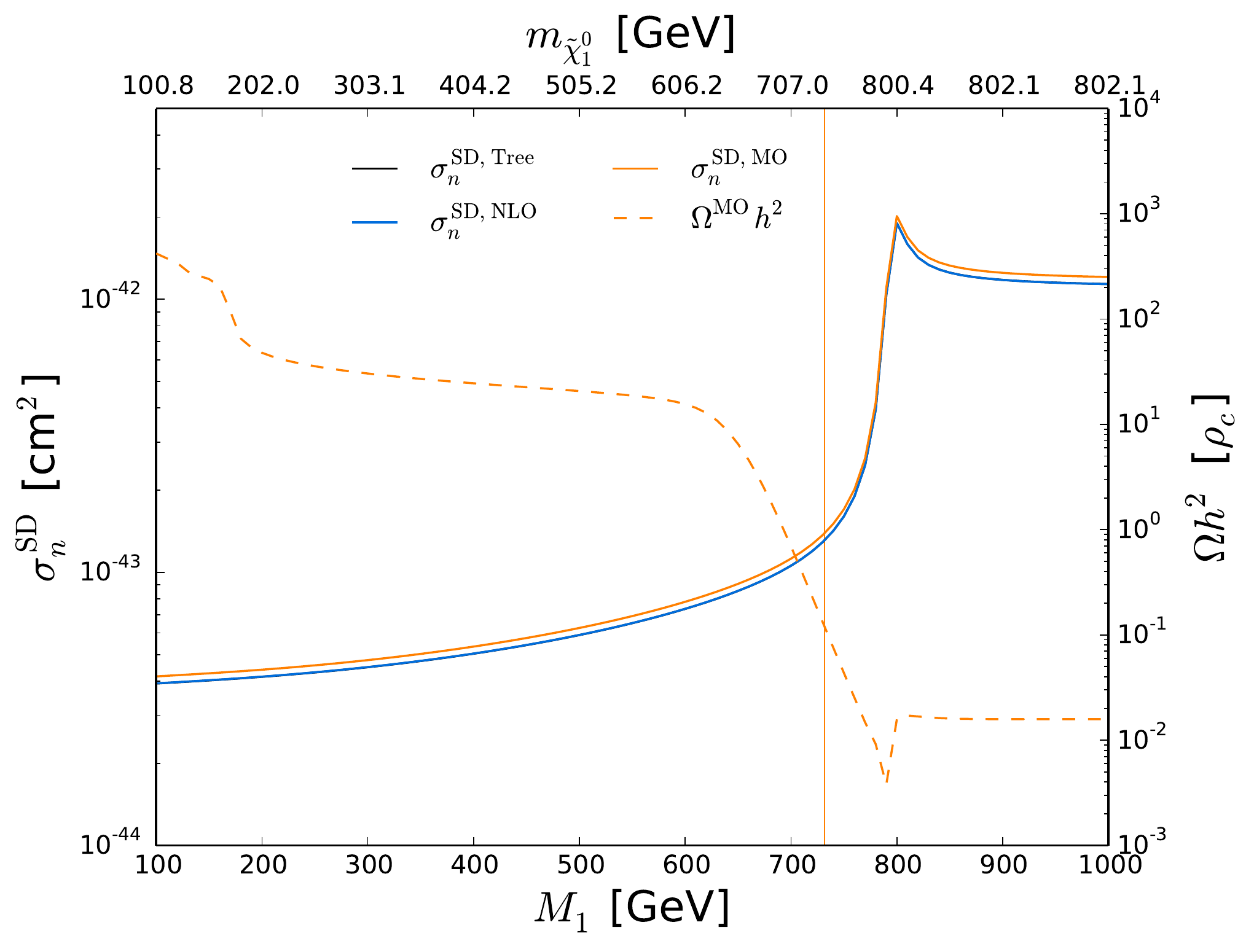}
	\caption{Spin-independent (top) and spin-dependent (bottom) neutralino-nucleon cross sections in scenario A for protons (left) and neutrons (right).}
	\label{fig:CrossSectionsScenA}
\end{figure*}

We continue with the discussion of the neutralino-nucleon cross sections, which are displayed in Fig.\ \ref{fig:CrossSectionsScenA}. The upper left plot of Fig.\ \ref{fig:CrossSectionsScenA} illustrates the spin-independent neutralino-proton cross section. This quantity has been calculated by \MO\ (orange solid line), our code at tree level (black solid line) and our code including full $\mathcal{O}(\alpha_s)$ corrections to the dominant effective operators (blue solid line). The shift between our tree-level calculation and \MO\ is mainly due to different nuclear input values (cf.\ Tab.\ \ref{fTTable}). After adjusting the nuclear input, our tree-level calculation agrees quite well with \MO, which is shown by the dotted black line. In absolute numbers, as expected, the neutralino-proton cross section is rather small (10$^{-47}$ - 10$^{-46}$ cm$^{2}$), as long as the neutralino is mostly bino. The tree-level couplings to Higgs bosons are supressed in this case, and so are the squark processes because of the heavy squark masses (cf.\ Tab.\ \ref{ScenarioProps}). The shift between our tree-level and our full NLO calculation is of similar size as the shift between our tree level and \MO. In the present case, the first shift is mainly caused by SUSY-QCD corrections to the Higgs exchange process including third generation squarks as the other squarks are much heavier.

Furthermore we show the improved\footnote{More precisely, the green dotted line corresponds to the choice \texttt{MSSMDDTest(loop=1, ...)}, whereas the orange solid line corresponds to \texttt{MSSMDDTest(loop=0, ...)}.} tree-level calculation of \MO\ as the green dotted line. Among other improvements, this choice is supposed to replace the heavy quark contributions by the gluon one-loop processes as given in Ref.\ \cite{Drees}. However, we could not find a significant difference in comparison to the pure tree-level calculation in any scenario. Therefore, the green dotted and orange full lines are indistinguishable also in this plot.

We also show the resulting relic density obtained with \MO\ as the dashed orange line (right ordinate). Note that this curve is roughly inverse to the cross section curves. This correlation is not completely unexpected. Larger gaugino (co)annihilation cross sections into final quark states leading to a smaller relic density are linked to larger neutralino-nucleon cross sections. The crucial condition for this correlation is that the neutralinos annihilate dominantly into quark final states. In the present case this is given for $M_1 > 200$ GeV. For smaller $M_1$, neutralinos prefer to annihilate into electroweak final states, and the resulting bump in the relic density has no counterpart in the neutralino-nucleon cross section. The orange vertical band marks the region of $M_1$ leading to a relic density compatible with the Planck limits as given in Eq.\ (\ref{Planck}). We investigate this region in greater detail later.

The upper right plot of Fig.\ \ref{fig:CrossSectionsScenA} shows the spin-independent neutralino-neutron cross section. No major difference in comparison to the proton case is found in this scenario, since the isospin-dependent contributions from first-generation quarks are suppressed by large squark masses.

We continue with the lower left plot of Fig.\ \ref{fig:CrossSectionsScenA} where the spin-dependent neutralino-proton cross section is given. Here the blue and black solid lines completely overlap, signalizing that the NLO corrections are negligible. This is indeed the case in this scenario. Remember that only light quarks ($u,d,s$) and corresponding squarks contribute to the spin-dependent cross section (cf.\ subsection \ref{DDFormulas}). These squarks are very heavy in this scenario (cf.\ Tab.\ \ref{ScenarioProps}) and loops including them are strongly suppressed. The small shift ($\sim + 7\%$) between our results and \MO\ is not due to the nuclear input values this time -- by default we are using the same input in the spin-dependent case. It is rather due to the running of the operator and associated Wilson coefficient described in subsection \ref{RunningSection}, which is not implemented in \MO. If we deactivate the running in our code, we find perfect agreement with \MO. The spin-dependent neutralino-neutron cross section is shown in the lower right plot of Fig.\ \ref{fig:CrossSectionsScenA}. As before, no major difference in comparison to the proton case is found in this scenario.

\begin{figure}
	\includegraphics[width=0.49\textwidth]{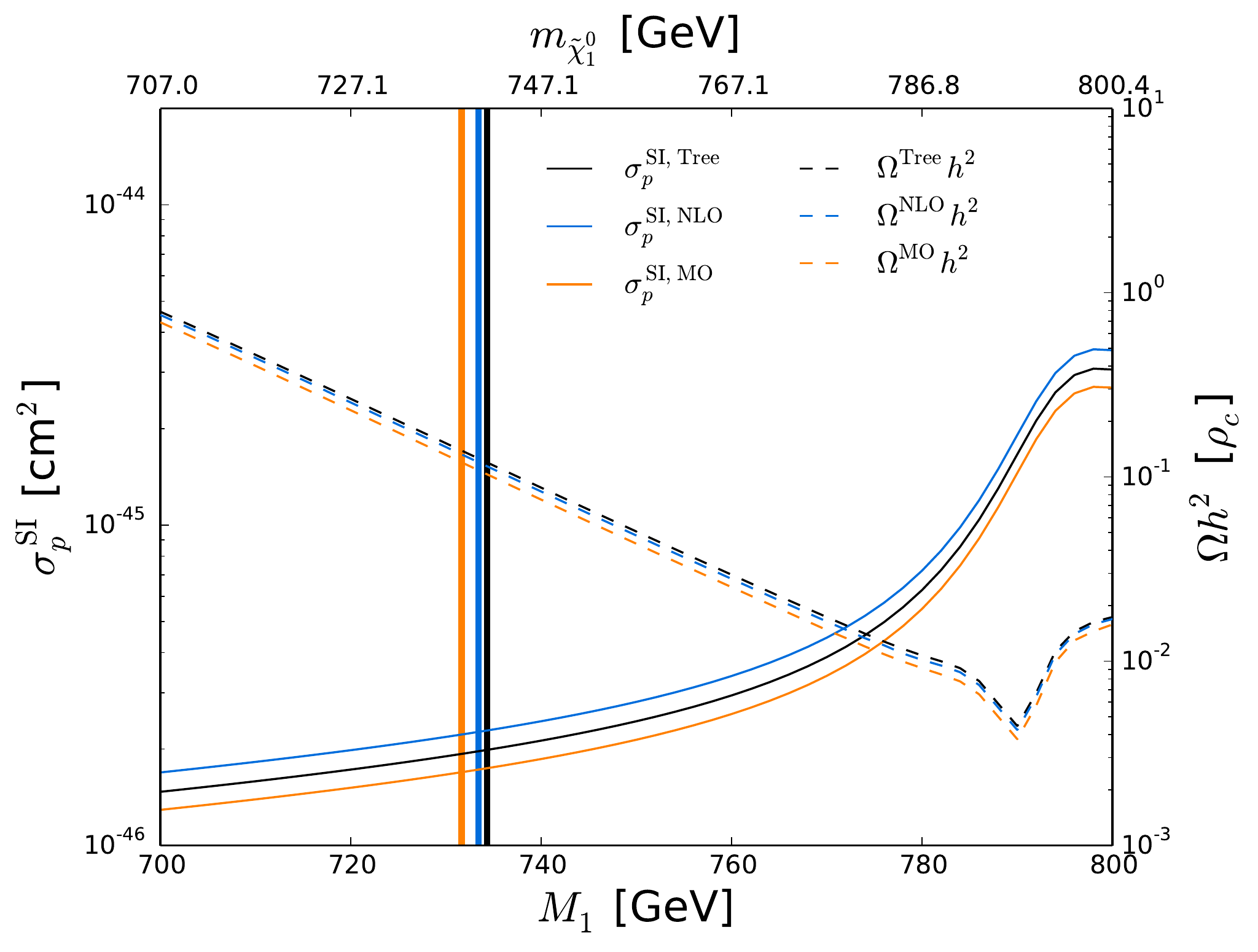}
	\caption{Combined relic density and direct detection calculation in scnenario A.}
	\label{fig:ZoomScenA}
\end{figure}

We take a closer look at the cosmologically preferred region now, i.e.\ we zoom into the region 700 GeV $< M_1 <$ 800 GeV of the upper left plot of Fig.\ \ref{fig:CrossSectionsScenA}. The result is shown in Fig.\ \ref{fig:ZoomScenA}. Apart from the previously introduced three solid lines, we depict the relic density obtained with \MO\ (orange dashed line), our code at tree level (black dashed line) and our code at NLO (blue dashed line). These three calculations lead to different cosmologically preferred regions as indicated by the orange, black and blue vertical band, respectively. Assuming that the neutralinos solely account for dark matter, we can combine these calculations to constrain the pMSSM parameter space and to precisely predict the resulting neutralino-nucleon cross section. This corresponds to identifying the intersections of the vertical bands and solid lines of the same color. The results are given in Tab.\ \ref{PredictionsScenA} where we also list the relative shifts of the \MO\ and our full NLO result with respect to our tree-level calculation. The shifts are in opposite directions and of similar size in this case.

\begin{table}
	\caption{Resulting $M_1$ and spin-independent neutralino-proton cross section when combining direct detection and relic density routines in scenario A.}
	\begin{tabular}{|c|ccc|}
		\hline
			$\quad$ & $M_1$ [GeV] & $\sigma^{\mathrm{SI}}_p$ [$10^{-46}$cm$^2$]& Shift of $\sigma^{\mathrm{SI}}_p$\\ 
			\hline 
			\MO\ & 731 & $1.68$ & $-15\%$ \\
			Tree level & 734 & $1.98$ &  \\	
			Full NLO &  733 & $2.26$ & $+14\%$ \\			
			\hline
	\end{tabular}
	\label{PredictionsScenA}
\end{table}

\subsection*{Scenario B -- Bino-higgsino dark matter}

\begin{figure}
	\includegraphics[width=0.49\textwidth]{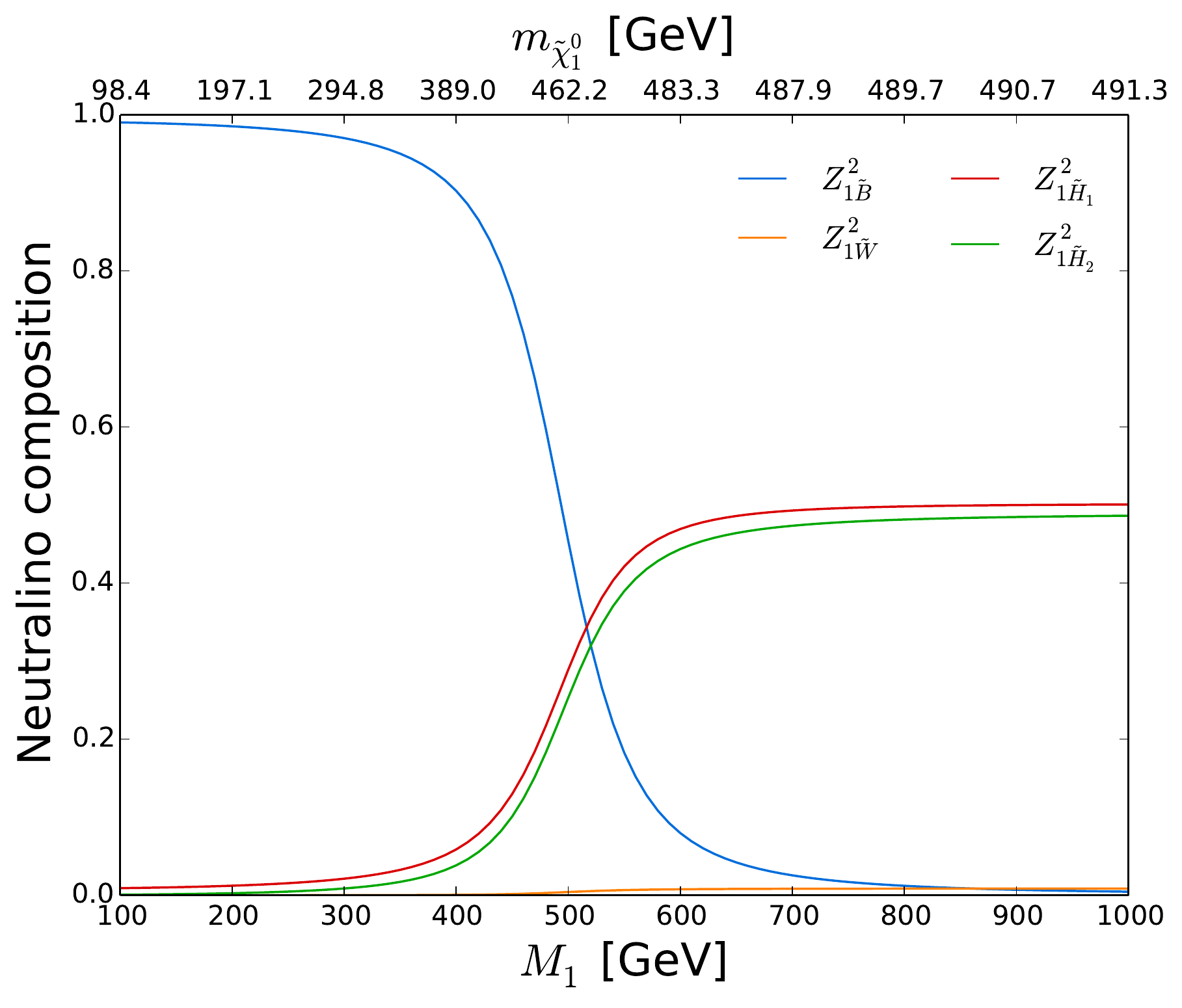}
	\caption{Neutralino decomposition in scenario B.}
	\label{fig:MixingScenB}
\end{figure}

When varying $M_1$ in scenario B, the neutralino decomposition changes again, this time from mostly bino into mostly higgsino as shown in Fig.\ \ref{fig:MixingScenB}. The turning point is at $M_1 \sim \mu \sim 500$ GeV. The neutralino mass depends only weakly on $M_1$ for larger values of $M_1$. In comparison to the previous scenario, the remaining dependence is larger which is in agreement with the softer admixture transition (compare Figs.\ \ref{fig:MixingScenA} and \ref{fig:MixingScenB}). This decomposition and the relatively light squarks (cf.\ Tab.\ \ref{ScenarioProps}) are the essential phenomenological properties of this scenario.

\begin{figure*}
	\includegraphics[width=0.49\textwidth]{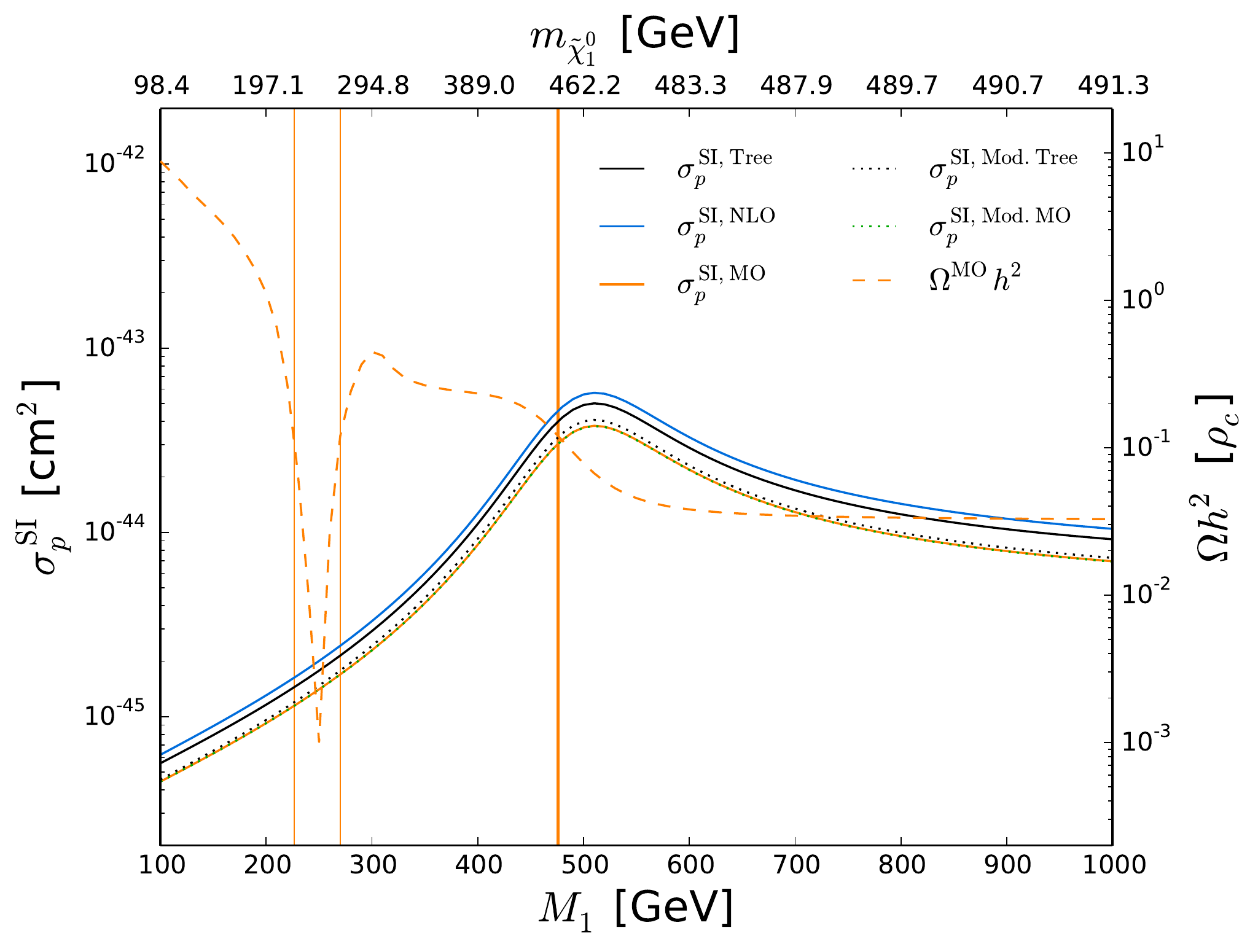}
	\includegraphics[width=0.49\textwidth]{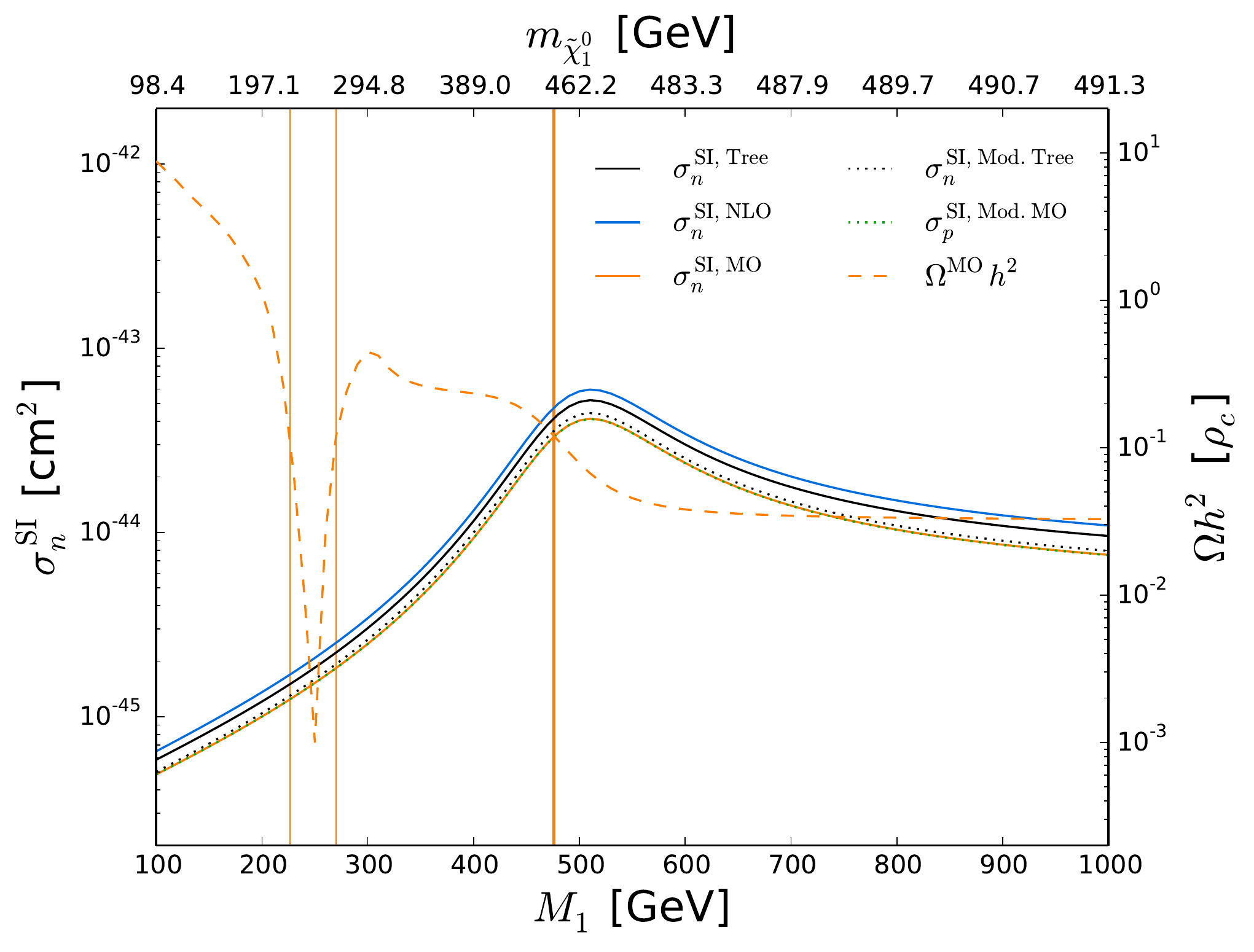}
	\includegraphics[width=0.49\textwidth]{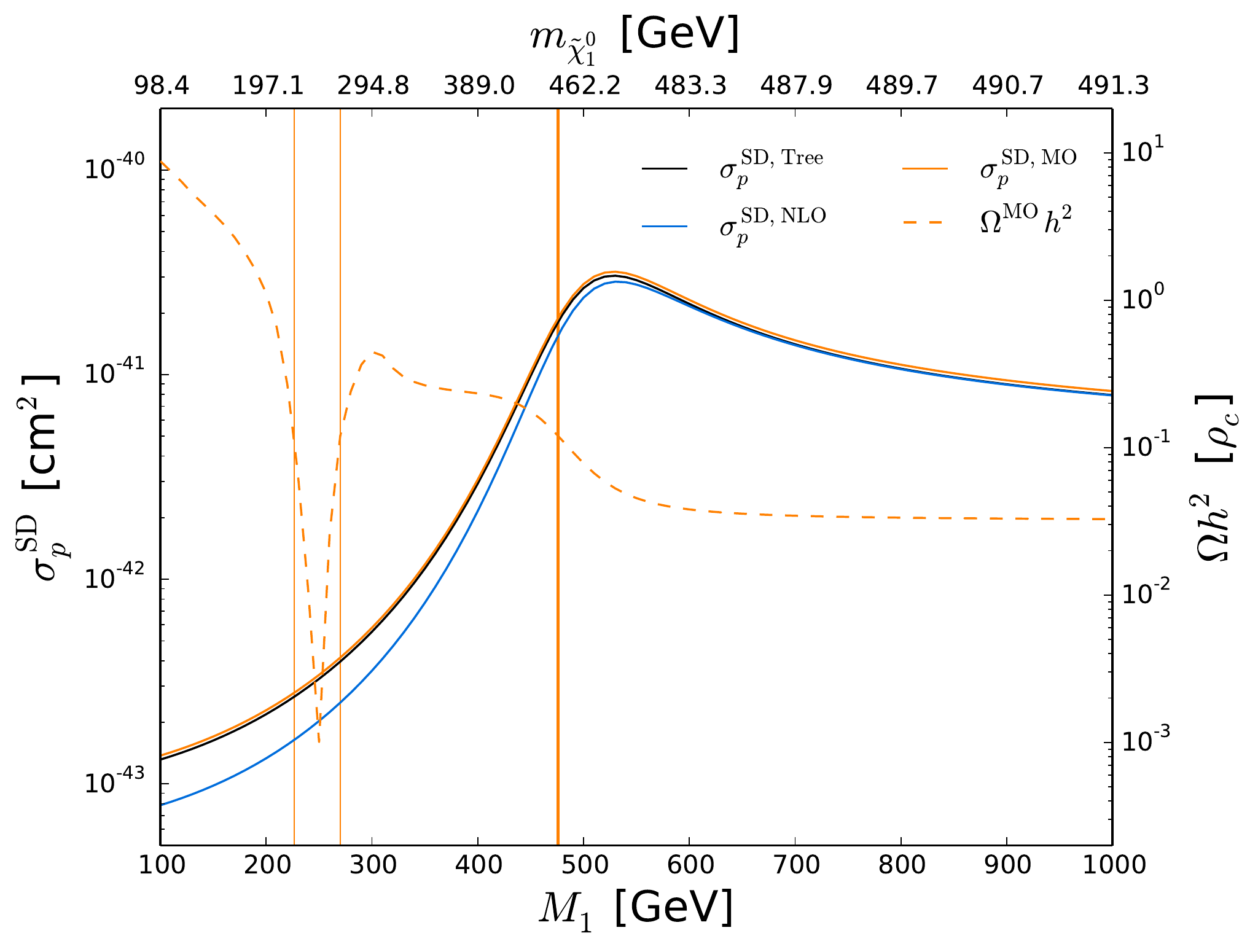}
	\includegraphics[width=0.49\textwidth]{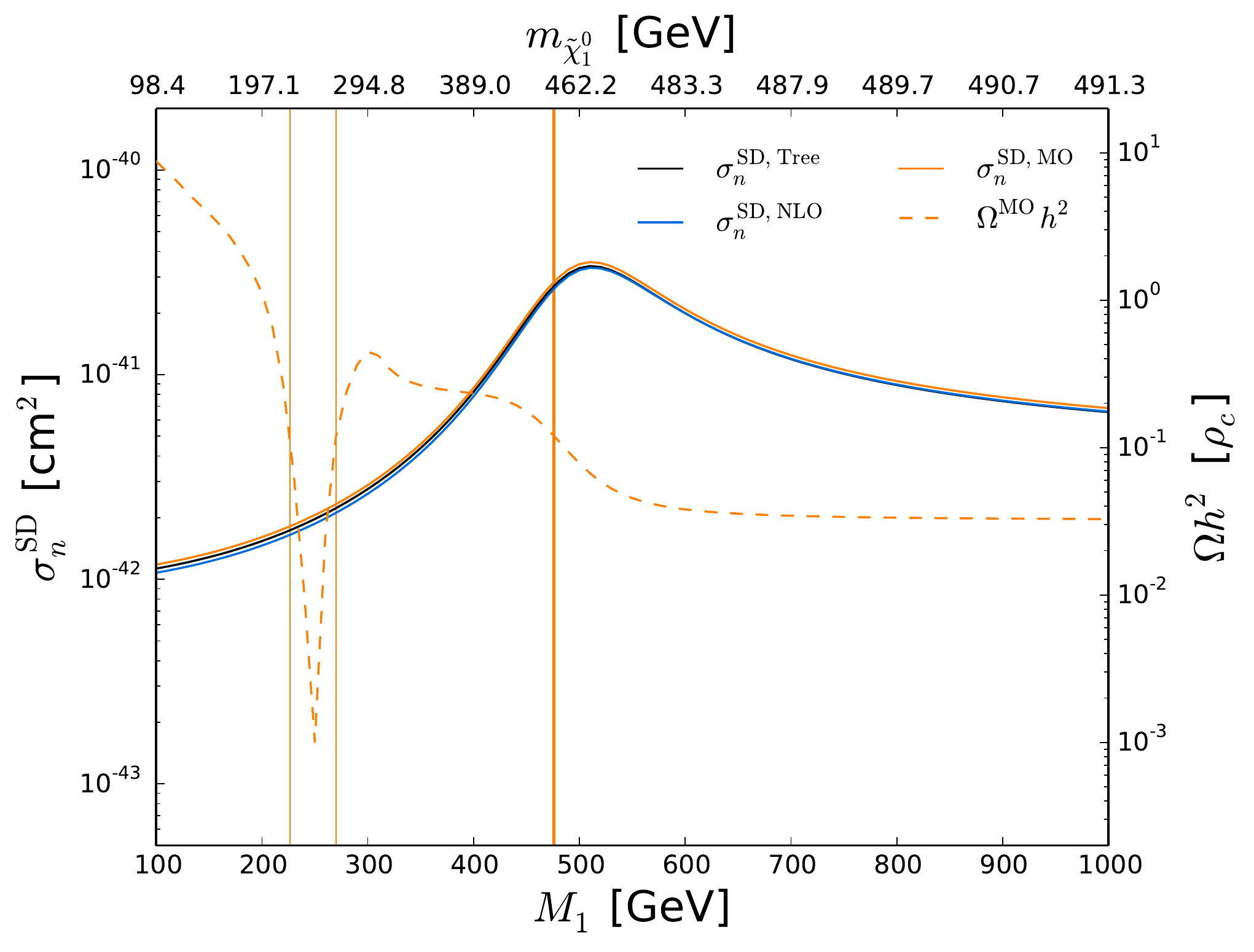}
	\caption{Spin-independent (top) and spin-dependent (bottom) neutralino-nucleon cross sections in scenario B for protons (left) and neutrons (right).}
	\label{fig:CrossSectionsScenB}
\end{figure*}

The neutralino-nucleon cross sections for scenario B are shown in Fig.\ \ref{fig:CrossSectionsScenB}. The first thing to note is that there are three vertical orange bands now, corresponding to three regions which lead to a relic density compatible with Eq.\ (\ref{Planck}). Apart from scenario B itself ($M_1 = 270$ GeV), there is a second line on the other side of the peak of the dashed orange line and a third one at $M_1 \sim 475$ GeV. The peak is due to a Higgs resonance caused by $2m_{\tilde{\chi}^0_1} \sim m_{H^0}, m_{A^0} \sim 500$ GeV, which heavily increases the neutralino cross section into bottom quarks and in turn heavily reduces the resulting relic density. Bottom quarks are favored over top quarks, as $\tan\beta = 13.7$ is rather large here. The peak does not show up in the neutralino-nucleon cross sections. This is as expected, as the Higgs process has turned from a resonant $s$-channel to a non-resonant $t$-channel. The third vertical band lies precisely in the region where the neutralino admixture changes from bino to higgsino, stressing again the necessity to treat the general neutralino admixture.

The spin-independent nucleon cross sections are shown in the upper plots of Fig.\ \ref{fig:CrossSectionsScenB}. Once again, no major difference is found between the proton and neutron case. The relative shifts between our tree-level calculation (black solid line) and \MO\ (orange solid line) or our NLO calculation (blue solid line) are roughly as before. No significant change is found when activating the improved tree-level calculation of \MO\ (green dotted line). The agreement between our tree-level calculation using the nuclear input values of \MO\ (black dotted line) and the \MO\ result is slightly worse. The remaining discrepancy is mainly due to the use of effective couplings in \MO\ and a different treatment of the top quark mass and of the associated stop sector (cf.\ subsection \ref{Renormalization}). Moreover \MO\ does not kinematically distinguish between the $s$- and the $u$-channels shown in Fig.\ \ref{fig:DDTree}. Although these differences are present in general, the resulting discrepancy depends on the concrete scenario. In this scenario they lead to a small, but visible shift, whereas they do not in the other two scenarios.

New features show up in the spin-dependent case, i.e.\ in the lower plots of Fig.\ \ref{fig:CrossSectionsScenB}. Here, the proton and neutron cross sections differ by almost one order of magnitude in the small $M_1$ regime. Moreover the tree-level and NLO results clearly separate in the proton case for small $M_1$. This large splitting is absent in the neutron case. The reason is as follows: In the small $M_1$ regime, the neutralino is mostly bino (cf.\ Fig.\ \ref{fig:MixingScenA}). Moreover, the squarks of the first two generations are rather light in this scenario (cf.\ Tab.\ \ref{ScenarioProps}). The former leads to a suppression of the usually dominant $Z^0$ processes, while the latter kinematically favors the squark processes. As a result, the squark processes contribute sizeably in the small $M_1$ regime. In contrast to the $Z^0$ processes, these processes strongly depend on the involved quark flavor and are sensitive to different choices of $(\Delta_q)_N$ as given in Eqs.\ (\ref{Delta1}) and (\ref{Delta2}). In the case of the proton, this leads to a partial cancellation of the individually large squark contributions, which is much less pronounced in the neutron case. This explains the difference between the proton and neutron cross sections. The rather large impact of the NLO corrections on the proton cross section has a related origin. As the leading squark contributions cancel in the proton case, the cross section becomes more sensitive to the subleading virtual corrections. Due to the rather light squark masses in this scenario, these virtual corrections are not negligible. For large $M_1$, the $Z^0$ processes dominate and the virtual corrections are less important.

\begin{figure}
	\includegraphics[width=0.49\textwidth]{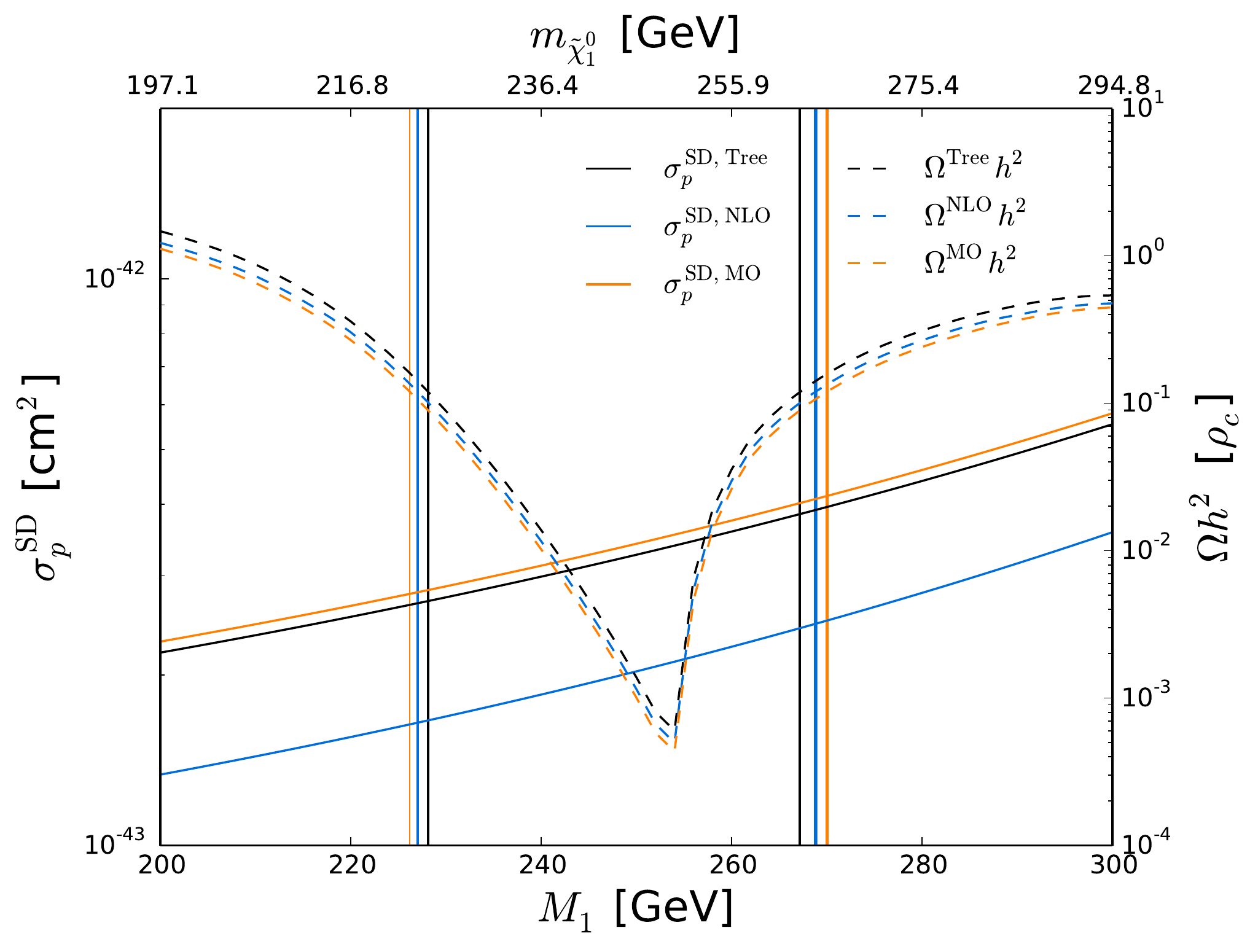}
	\caption{Combined relic density and direct detection calculation in scnenario B.}
	\label{fig:ZoomScenB}
\end{figure}

We take a closer look at the cosmologically preferred region around the Higgs resonance in the case of the spin-dependent neutralino-proton cross section in Fig.\ \ref{fig:ZoomScenB}. As before, we are showing the resulting relic density obtained with \MO\ (orange dashed line), our tree-level calculation (black dashed line) and our NLO calculation (blue dashed line). The vertical bands of the respective colors correspond to the $M_1$ regions leading to a relic density compatible with Eq.\ (\ref{Planck}). These bands are very thin here, as the relic density is changing rapidly near the resonance, which allows to effectively constrain the pMSSM parameter space. Subsequently we can read off the predicted cross section. The results are shown in Tab.\ \ref{PredictionsScenB}.

\begin{table}
	\caption{Resulting $M_1$ and spin-dependent neutralino-proton cross section when combining direct detection and relic density routines in scenario B.}
	\begin{tabular}{|c|ccc|}
		\hline
			$\quad$ & $M_1$ [GeV] & $\sigma^{\mathrm{SD}}_p$ [$10^{-43}$cm$^2$]& Shift of $\sigma^{\mathrm{SD}}_p$\\ 
			\hline 
			\MO\ & 226 & $2.78$ & $+3\%$ \\
			Tree level & 228 & $2.70$ &  \\	
			Full NLO &  227 & $1.65$ & $-39\%$ \\	
			\hline
			\MO\ & 270 & $4.14$ & $+8\%$\\
			Tree level & 267 & $3.84$ &  \\	
			Full NLO & 269 & $2.47$ & $-36\%$ \\				
			\hline
	\end{tabular}
	\label{PredictionsScenB}
\end{table}

As we are using the same nuclear input as \MO\ in the spin-dependent case, the shift between our tree-level prediction and \MO\ is smaller than in scenario A, where we investigated the spin-independent neutralino-proton cross section. Note that the relative position of the vertical bands, i.e.\ the relic density constraint, can influence this shift in both directions. The effect of reading off the cross section at different $M_1$ reduces the shift in the first case ($M_1= 226$ GeV and $M_1 = 228$ GeV) and increases the shift in the second case ($M_1 = 270$ GeV and $M_1 = 267$ GeV), as the order of the bands has changed. The exact opposite occurs when comparing our tree-level and our NLO results. Here both relative shifts are large, reaching almost $-40\%$.

\subsection*{Scenario C -- Higgsino-bino dark matter}
\label{ScenarioC}

\begin{figure}
	\includegraphics[width=0.49\textwidth]{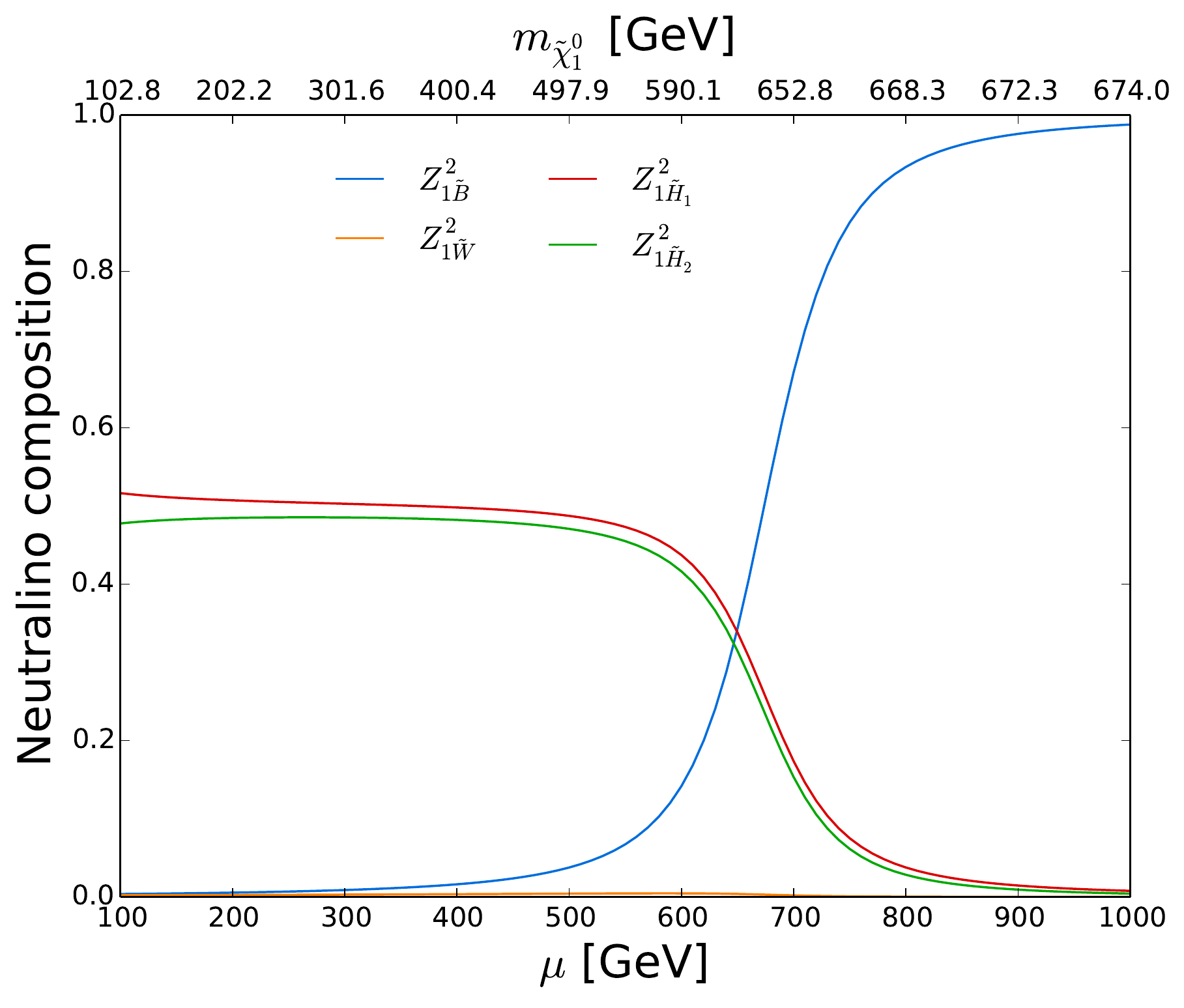}
	\caption{Neutralino decomposition in scenario C.}
	\label{fig:MixingScenC}
\end{figure}

In scenario C, we vary the higgsino mass parameter $\mu$, which changes the neutralino decomposition from higgs\-ino to bino as shown in Fig.\ \ref{fig:MixingScenC}. The turning point is at $\mu \sim M_1 \sim 675$ GeV. Concerning the neutralino admixture, scenario C can be understood as a mirrored version of scenario B (cf.\ Fig.\ \ref{fig:MixingScenB}).

\begin{figure*}
	\includegraphics[width=0.49\textwidth]{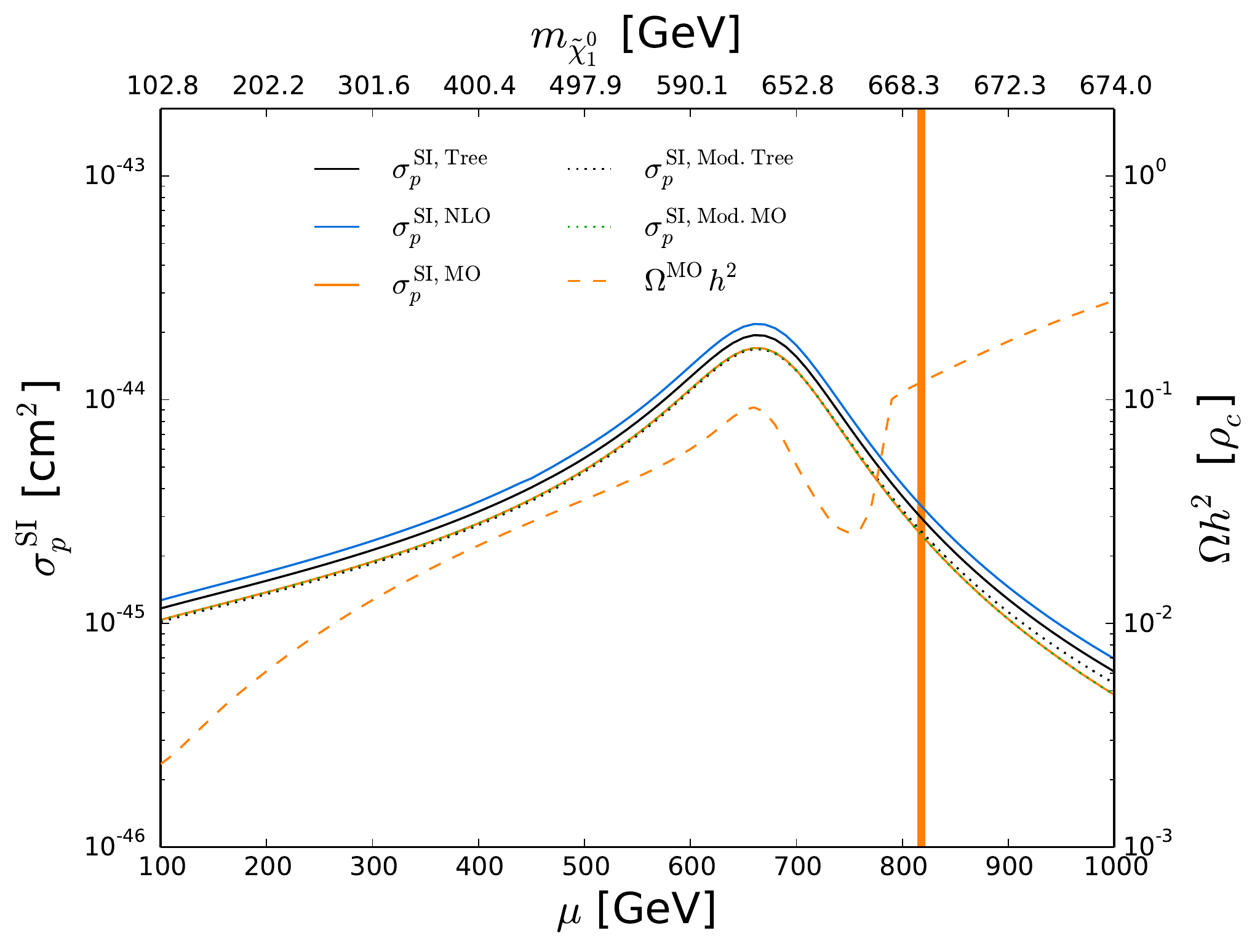}
	\includegraphics[width=0.49\textwidth]{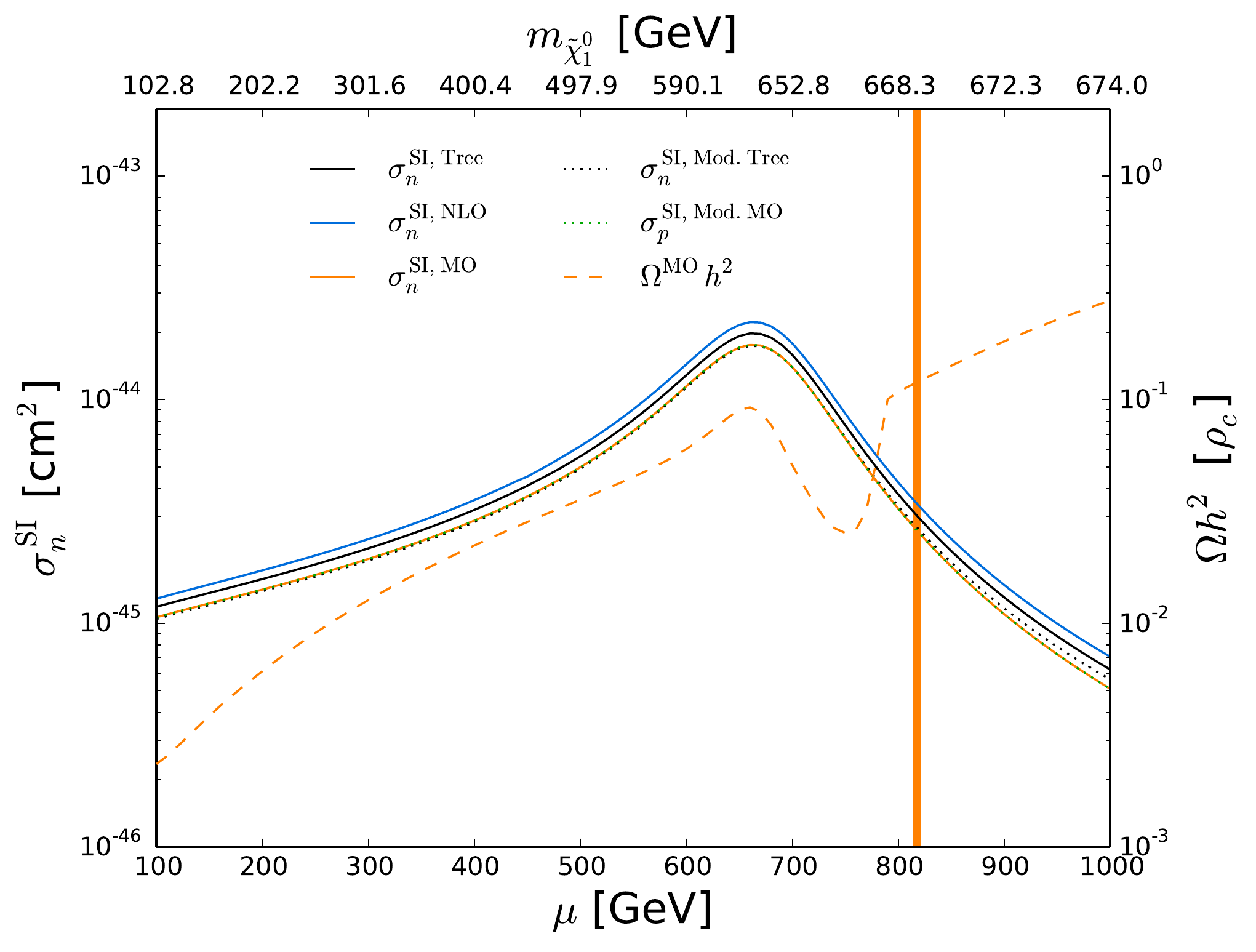}
	\includegraphics[width=0.49\textwidth]{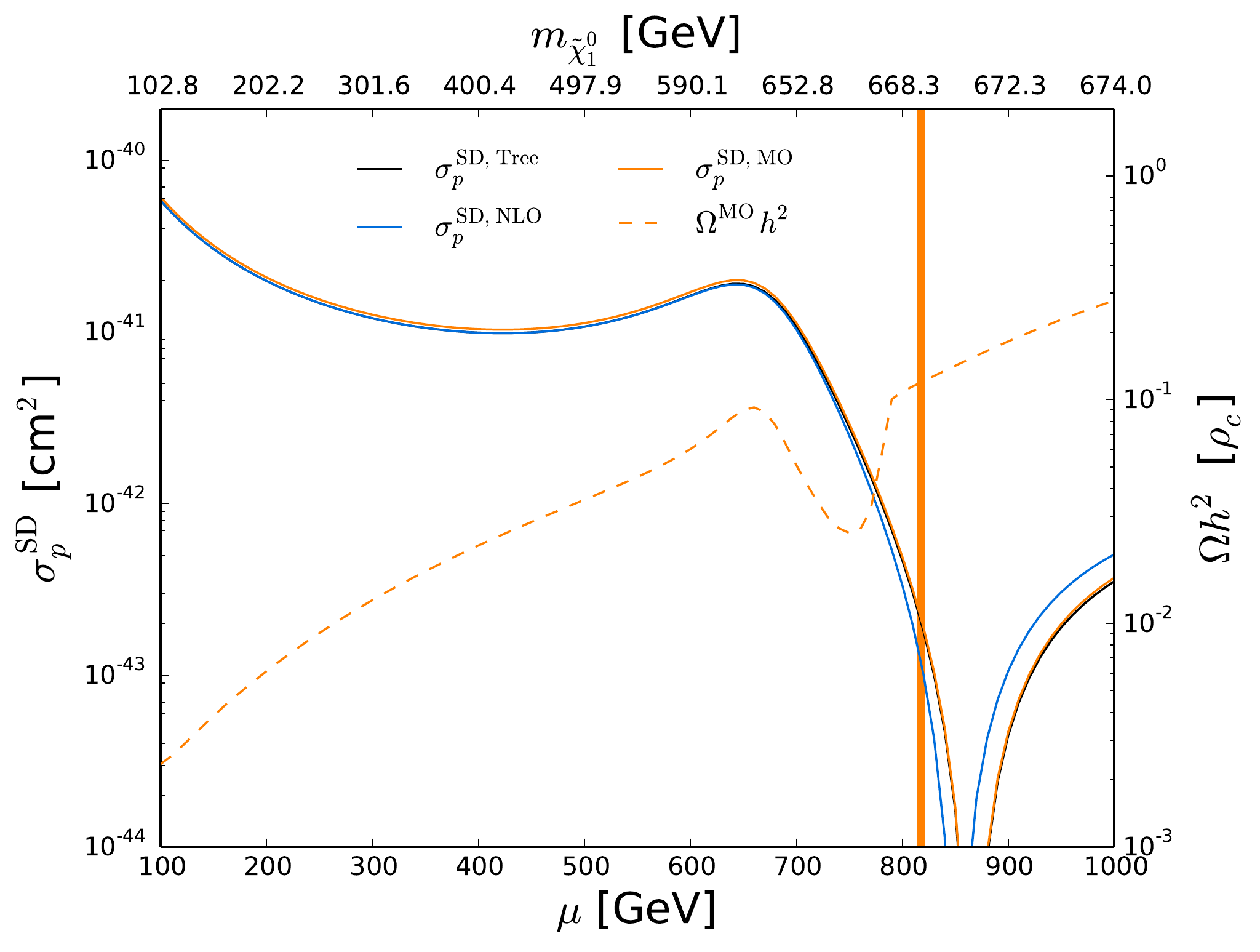}
	\includegraphics[width=0.49\textwidth]{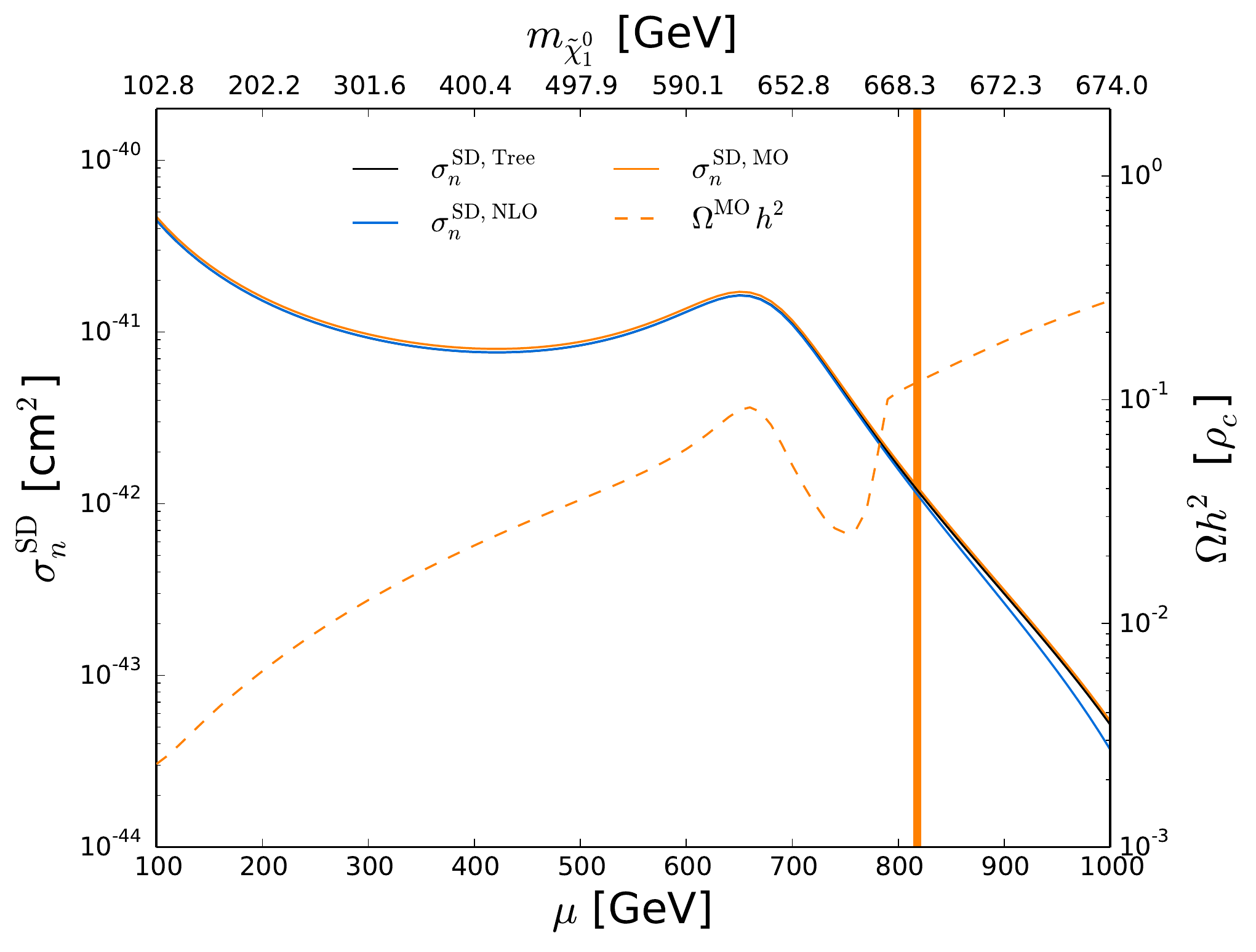}
	\caption{Spin-independent (top) and spin-dependent (bottom) neutralino-nucleon cross sections in scenario C for protons (left) and neutrons (right).}
	\label{fig:CrossSectionsScenC}
\end{figure*}

The neutralino-nucleon cross sections in scenario C are shown in Fig.\ \ref{fig:CrossSectionsScenC}. As the slope of the relic density (orange dashed line) is smaller than in the previous scenarios, the region compatible with the Planck limits is larger, which leads to a thicker vertical orange band. No essential new features are found in the spin-independent cross sections shown in the upper plots of Fig.\ \ref{fig:CrossSectionsScenC}. In particular, the relative shift between our tree-level calculation (black solid line) and \MO\ (orange solid line) for a given $M_1$ is roughly as big as the shift between our tree-level and our NLO calculation (blue solid line) amounting to $\sim-16\%$ and $\sim+13\%$, respectively. No significant difference is found between the proton and the neutron case. 

In contrast to that, the spin-dependent cross sections shown in the lower plots of Fig.\ \ref{fig:CrossSectionsScenC} obviously depend on the nucleon type. This difference is caused by a similar phenomenon as the one described in the previous subsection. In the large $\mu$ region, the neutralino becomes mostly bino (cf.\ Fig.\ \ref{fig:MixingScenC}), which suppresses the $Z^0$ processes. Although the squarks are not as light as in scenario B here (cf.\ Tab.\ \ref{ScenarioProps}), the squark processes are kinematically favored again. Remember that the squark processes occur in the $s$- and $u$-channel. The denominators of the tree-level processes read $s/u - m^2_{\tilde{q}_i}$ which simplifies to $(m_{\tilde{\chi}^0_1} \pm m_q)^2 - m^2_{\tilde{q}_i}$ in the limit of vanishing relative velocity. Hence it is not the total squark mass, but the neutralino-squark mass difference that matters. This difference decreases with increasing $\mu$. As a result, the squark processes contribute sizeably to the spin-dependent cross sections for large $\mu$. These processes depend on the involved flavor and in turn the chosen nuclear input values as given in Eqs.\ (\ref{Delta1}) and (\ref{Delta2}). In the case of the proton, we encounter a destructive interference of the individual terms, which leads to the drop observed at $\mu\sim 850$ GeV. Here the associated four-fermion coupling changes its sign and the resulting cross section vanishes. A similar situation would be encountered in the neutron case for larger values of $\mu$. However, as this region leads to a too large relic density, we are not investigating this in more detail.

\begin{figure}
	\includegraphics[width=0.49\textwidth]{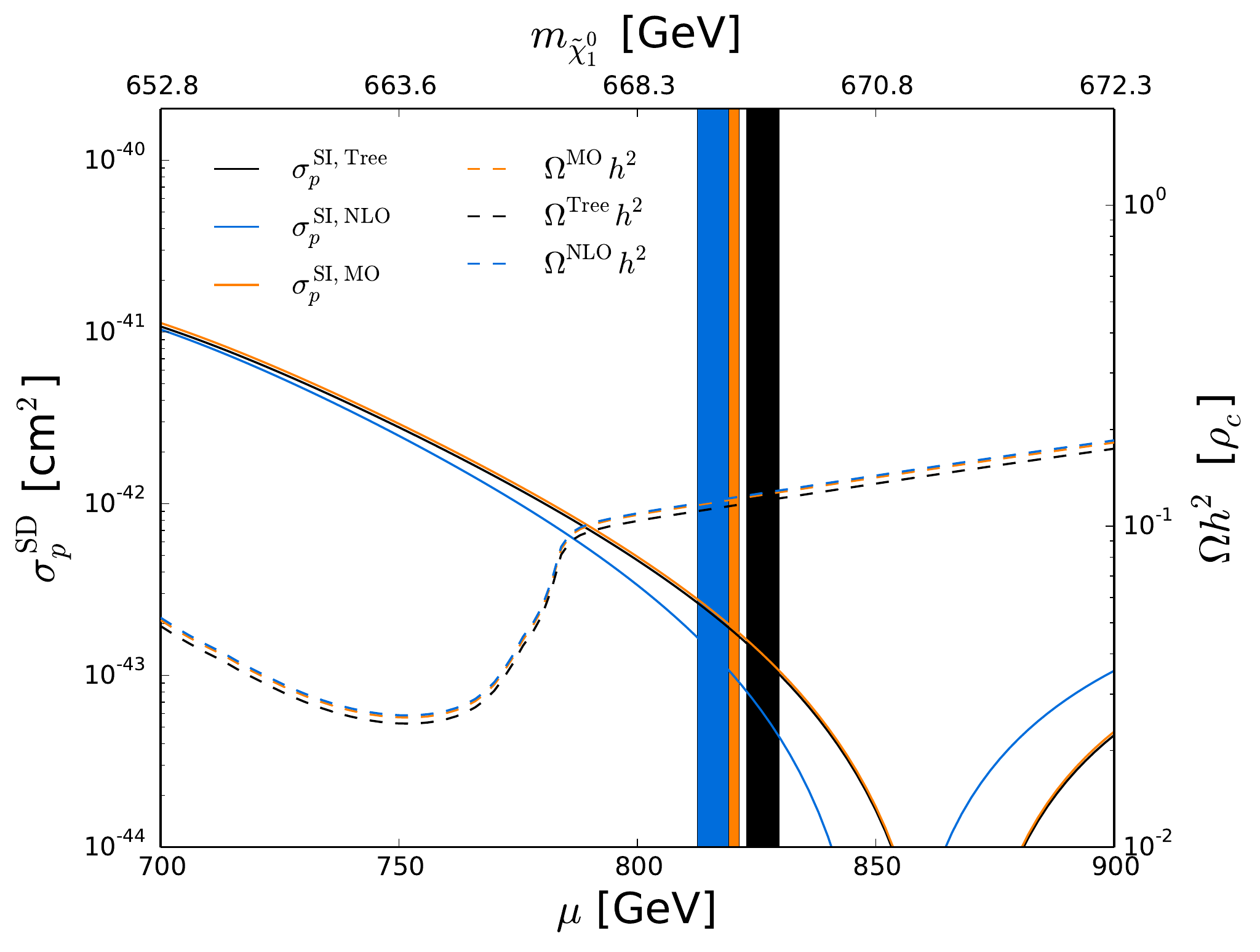}
	\caption{Combined relic density and direct detection calculation in scnenario C.}
	\label{fig:ZoomScenC}
\end{figure}

Instead, we zoom into the region $700 < \mu < 900$ GeV and analyze the spin-dependent neutralino-proton cross section in Fig.\ \ref{fig:ZoomScenC}. As before, we are showing the resulting relic density obtained with \MO\ (orange dashed line), our tree-level (black dashed line) and our NLO routines (blue dashed line). These three calculations lead to different regions compatible with the Planck limits, as indicated by the vertical bands of the corresponding colors. The bands are broader than before, as the relic density is increasing less rapidly when changing $\mu$. Note that the blue and orange bands overlap to a large extent, which signals that the effective couplings used in the relic routines of \MO\ are able to approximate the dominant NLO contributions quite well in this scenario. This may happen, but is not necessarily the case, as studied e.g. in Ref. \cite{ChiChi2qq3}. On the other hand, \MO\ does not include radiative corrections to the spin-dependent cross section. Hence the orange solid line is closely following the black solid line. As mentioned before, the remaining difference is due to the running of the operator and associated Wilson coefficient.

It is again interesting to combine the relic density and direct detection calculations. Our tree-level and NLO routines lead to different preferred regions along the $\mu$ axis. Simultaneosly the shift between the cross section obtained at tree level and at NLO is very large for a given $\mu$ (more than $-50\%$ near the drop). However, when combining both calculations these effects cancel each other. This is not the case for the comparison of \MO\ with our tree-level result, where \MO\ predicts a larger cross section.

The aforementioned regions of $\mu$ and the corresponding cross sections are listed in Tab.\ \ref{PredictionsScenC}. The broader vertical bands result in a range of allowed $\mu$ values and an associated range of cross sections. Note that these ranges exist in principle in every scenario. However, as they are very small in the previous scenarios we omitted them for simplicity. The shifts given in Tab.\ \ref{PredictionsScenC} are exemplary and have been obtained by combining the mean values of the cross sections. 

\begin{table}
	\caption{Resulting $\mu$ and spin-dependent neutralino-proton cross section when combining direct detection and relic density routines in scenario C.}
	\begin{tabular}{|c|ccc|}
		\hline
			$\quad$ & $\mu$ [GeV] & $\sigma^{\mathrm{SD}}_p$ [10$^{-43}$cm$^2$]& Shift of $\sigma^{\mathrm{SD}}_p$\\ 
			\hline 
			\MO\ & 815 - 821 & 1.80 - 2.43 & $+63\%$ \\
			Tree level & 823 - 829 & 1.06 - 1.53 &  \\	
			Full NLO &  813 - 819 & 1.08 - 1.62 & $+4\%$ \\			
			\hline
	\end{tabular}
	\label{PredictionsScenC}
\end{table}

\begin{figure}
	\includegraphics[width=0.49\textwidth]{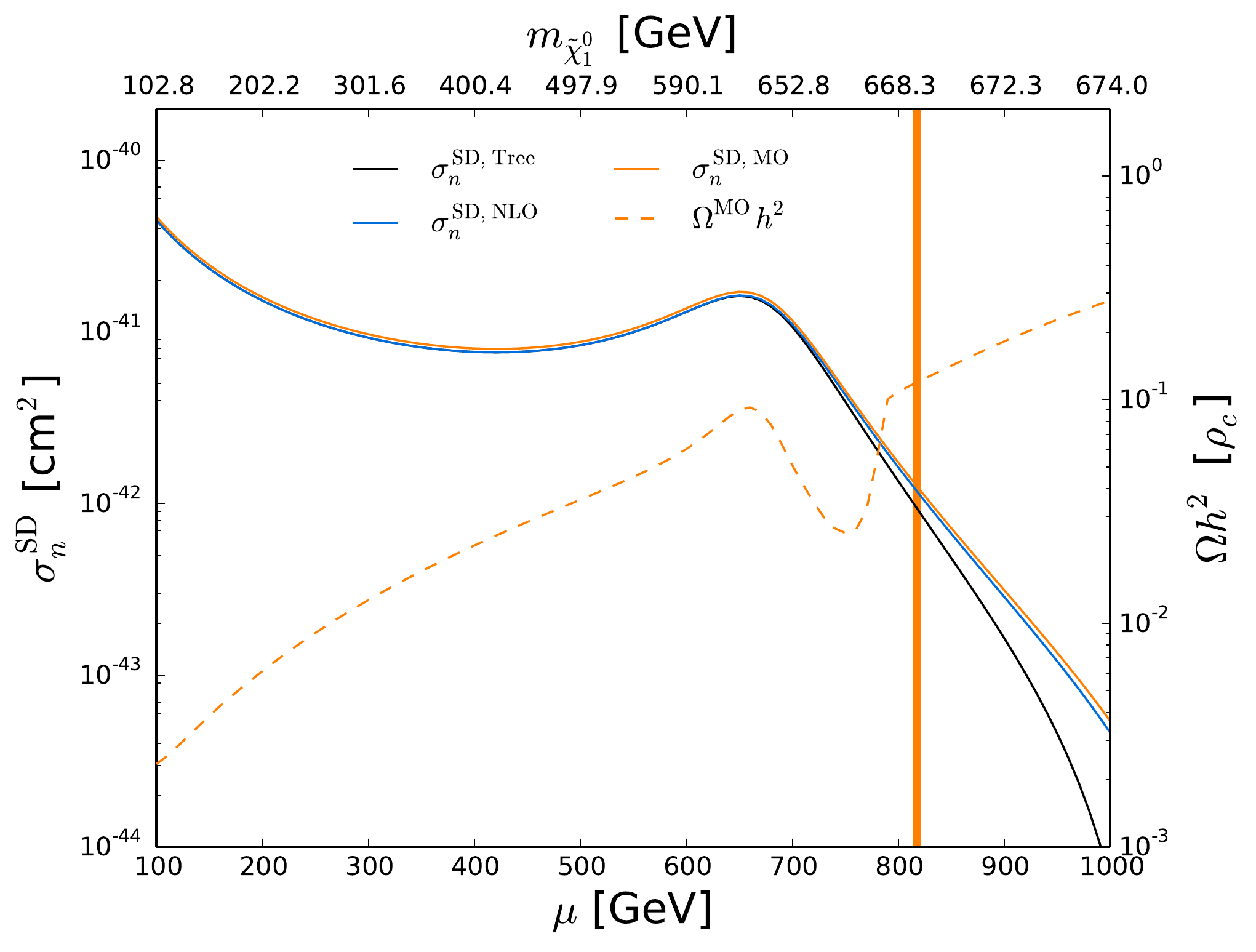}
	\caption{Spin-dependent neutralino-neutron cross section in scenario C using a pure \DRbar\ scheme}
	\label{fig:DRbar}
\end{figure}

Before concluding, we take a small detour and briefly comment on the renormalization scheme dependence. As we have described in subsection \ref{Renormalization}, we are working with a hybrid on-shell/\DRbar\ scheme. In particular the squark masses of the first two generations are treated on-shell just like in \MO. Our code optionally also supports a pure \DRbar\ scheme. When studying the differences between the two schemes, the spin-dependent neutralino-neutron cross section shown in the lower right plot of Fig.\ \ref{fig:CrossSectionsScenC} has proven very useful. This plot is shown again in Fig.\ \ref{fig:DRbar}, this time using a pure \DRbar\ scheme. No visible differences between the two plots occur for small $\mu$. The virtual corrections to the spin-dependent cross section are negligible in this regime which is not affected by the choice of the scheme. For larger values of $\mu$, our tree-level result (black solid line) -- now using \DRbar\ squark masses -- clearly separates from \MO\ (orange solid line) -- still using on-shell squark masses -- which has previously not been the case (cf.\ Fig.\ \ref{fig:CrossSectionsScenC}).

Remember that the cross section in this region is heavily influenced by the squark processes as explained in the beginning of this subsection. These processes benefit from the decreasing neutralino-squark mass difference appearing in the denominators of the corresponding propagators. This mass difference is sensitive to the choice of the scheme. To investigate this in greater detail, we write the scale-independent on-shell squark mass $m_{\tilde{q}}^{\mathrm{OS}}$ as a sum of two individually scale-dependent terms, the scale-dependent \DRbar\ mass $m_{\tilde{q}}^{\mathrm{\overline{DR}}}(\mu_R)$ and an additional finite term resumming virtual corrections $\Delta m_{\tilde{q}}(\mu_R)$,
\beq
m_{\tilde{q}}^{\mathrm{OS}} = m_{\tilde{q}}^{\mathrm{\overline{DR}}}(\mu_R) + \Delta m_{\tilde{q}}(\mu_R).
\eeq
If we replace the on-shell squark masses by their smaller \DRbar\ masses, i.e.\ if we discard the finite $\Delta m_{\tilde{q}}(\mu_R)$ terms, the neutralino-squark mass difference decreases even further. This leads to the observed steep drop of our tree-level result. However, at NLO this effect diminishes again and the blue lines shown in the lower right plot of Fig.\ \ref{fig:CrossSectionsScenC} and in Fig.\ \ref{fig:DRbar} roughly agree. The reason behind that is that the leading corrections incorporated in $\Delta m_{\tilde{q}}(\mu_R)$ reappear again, this time as virtual corrections to the squark propagators. In other words, the tree-level result heavily depends on the definition of the squark mass in this special situation, but the NLO result is much more stable. The main difference between the two schemes is that the virtual corrections are partially included at tree level in the on-shell mass in the first case, whereas they show up as large propagator corrections in the second case. We prefer the first scheme, which leads to smaller virtual corrections and an improved perturbative stability. Let us finally mention that the resulting differences between the two schemes in other cases, i.e.\ for other cross sections, are less pronounced. A similar study in the context of the relic density can be found in Ref.\ \cite{Scalepaper}.

%% file: conclusion.tex
\section{Conclusion}
\label{Conclusion}

In this paper, we presented a NLO SUSY-QCD calculation for the scattering of neutralino dark matter off of the partonic constituents of nucleons, which required a novel tensor reduction method of loop integrals with vanishing relative velocities and Gram determinants. We consistently matched these one-loop corrections to the scalar and axial-vector operators, which govern the spin-independent and spin-dependent scattering proceses in the effective field theory approach. As a result, the operators and Wilson coefficients aquired a scale dependence, which was taken into account by applying renormalization group running to the Wilson coefficients. Our formalism is valid for general compositions of bino, wino, or higgsino dark matter.

We investigated three benchmark scenarios, which satisfy current Higgs mass, relic density, flavor-changing neutral current and direct SUSY particle search constraints from the LHC, but which were not tuned to be particularly sensitive to the new NLO corrections for direct detection. Despite the fact that the first- and second generation squark masses were at the TeV scale, we observed corrections that were of similar size or in some cases larger than the currently estimated nuclear uncertainties. This could be explained by small neutralino-squark mass differences governing the propagator denominators at low velocity. In general, large corrections can be expected in the spin-independent case for Higgs bosons coupling to winos and heavy quarks, in the spin-dependent case for $Z$-bosons coupling to higgsinos and light (potentially also heavy) quark flavors, and in both cases from squarks with small masses or mass differences or scenarios with destructive interference at tree level. In the first case, our calculation is complementary to the explicit generation of heavy quarks from gluon operators at one loop, similarly to the complementarity of variable and fixed flavor schemes that are both employed in deep-inelastic scattering. The calculation for gluon operators has been performed previously elsewhere; its implementation in DM@NLO and a comparison of the two approaches is left for future work, as is a numerical study for light or nearly neutralino mass-degenerate squarks.

Through the implementation of direct detection at NLO as a second dark matter observable in DM@NLO, consistent investigations of correlations between direct detection and the relic density at NLO are now possible. First examples have been given in this paper in the three mentioned reference scenarios. Systematically, shifts in the extracted dark matter mass from NLO corrections to the relic density implied different NLO corrections to be expected in direct detection experiments.

%% file: appendix.tex
\section{Tensor reduction for vanishing Gram determinant}
\label{GramDet}

In the course of the \DMNLO\ project, we have computed a large collection of loop integrals and associated special cases in generic form. In addition, we always distinguish between infrared and ultraviolet divergences. As these divergences have to vanish when determining physical observables, this discrimination allows for powerful checks of our calculations. Hence it is desirable to use the same thoroughly tested routines for the new direct detection calculation. However, in the context of direct detection, all the amplitudes are evaluated at zero momentum transfer.\footnote{To determine the relic density, only cross sections including a finite relative velocity are needed, as those with zero relative velocity are weighted by zero in the thermal averaging procedure.} This causes problems for the tensor reduction of loop amplitudes which we employ \cite{PVIntegrals}.

In this appendix we present our alternative approach, which is partially based on Ref.\ \cite{Ganesh}. To keep the discussion transparent and to stress the general idea, we restrict ourselves to the simple case of determining the tensor coefficients $C_1$ and $C_2$. All other necessary tensor coefficients can be worked out analogously.

We start by setting up our notation. The scalar and tensor integrals relevant for our discussion are defined via 
\bea
B_0(p_1, m_0^2, m_1^2) & = & \frac{(2\pi\mu_R)^{4-D}}{i\pi^2}\int\mathrm{d}^Dq\frac{1}{\mathcal{D}_0\mathcal{D}_1},\label{B0}\nonumber\\
&&\\ 
B_\mu(p_1, m_0^2, m_1^2) & = & \frac{(2\pi\mu_R)^{4-D}}{i\pi^2}\int\mathrm{d}^Dq\frac{q_\mu}{\mathcal{D}_0\mathcal{D}_1},\label{Bmu}\nonumber\\
&&\\
C_0(p_1, p_2, m_0^2, m_1^2, m_2^2) & = & \frac{(2\pi\mu_R)^{4-D}}{i\pi^2}\int\mathrm{d}^Dq\frac{1}{\mathcal{D}_0\mathcal{D}_1\mathcal{D}_2},\label{C0}\nonumber\\
&&\\
C_\mu(p_1, p_2, m_0^2, m_1^2, m_2^2) & = & \frac{(2\pi\mu_R)^{4-D}}{i\pi^2}\int\mathrm{d}^Dq\frac{q_\mu}{\mathcal{D}_0\mathcal{D}_1\mathcal{D}_2},\label{Cmu}\nonumber\\
&&\\
C_{\mu\nu}(p_1, p_2, m_0^2, m_1^2, m_2^2) & = & \frac{(2\pi\mu_R)^{4-D}}{i\pi^2}\int\mathrm{d}^Dq\frac{q_\mu q_\nu}{\mathcal{D}_0\mathcal{D}_1\mathcal{D}_2}.\label{Cmunu}\nonumber\\
\eea
Here $\mu_R$ denotes the renormalisation scale which has been introduced to fix the mass dimension of the integrals. The denominators are given by $\mathcal{D}_i = (q+p_i)^2 - m_i^2 + i\epsilon$ with $p_0 = 0$. The idea of the tensor reduction method is to decompose the tensor integrals into a linear combination of all possible Lorentz structures accompanied by yet unknown tensor coefficients. Omitting the arguments, we have
\bea
B_\mu & = & p_{1,\mu}B_1,\\
C_\mu & = & p_{1,\mu}C_1 + p_{2,\mu}C_2,\\
C_{\mu\nu} &= & g_{\mu\nu}C_{00} + p_{1,\mu}p_{1,\nu}C_{11} + p_{2,\mu}p_{2,\nu}C_{22} \nonumber\\
&& + (p_{1,\mu}p_{2,\nu} + p_{2,\mu}p_{1,\nu})C_{12}.\label{StandardDecomposition}
\eea
The tensor coefficents are obtained by multiplying both sides with the available Lorentz invariants. In this way, the tensor integrals are reduced to a combination of scalar integrals. When determining $C_1$ and $C_2$ we have to solve a set of linear equations which results in
\bea
\begin{pmatrix} C_1 \\ C_2 \end{pmatrix} & = & A^{-1}\begin{pmatrix} R_1 \\ R_2 \end{pmatrix} =  \frac{1}{\det{A}}\begin{pmatrix} p_2^2 & -p_1p_2\\ -p_1p_2 & p_1^2\end{pmatrix}\begin{pmatrix} R_1 \\ R_2 \end{pmatrix}\label{C1-C2},\nonumber\\
&&
\eea
where we have introduced 
\bea
\det(A) & = & p_1^2p_2^2 -(p_1p_2)^2,\\
R_1  & = & \frac{1}{2}\left(B_0(0,2) - B_0(1,2) - f_1C_0\right),\label{R1}\\ 
R_2 & = &\frac{1}{2}\left(B_0(0,1) - B_0(1,2) - f_2C_0\right),\label{R2}\\
f_i & = & p_i^2 -m_i^2 + m_0^2\quad\mathrm{with}\quad i=1,2.\label{fi}
\eea
Furthermore we define the shorthand notation $B_0(i,j)  = B_0(p_j-p_i, m_i^2, m_j^2)$, which we use analogously for $B_1$.

The method illustrated above breaks down when the matrix $A$ is not invertible, i.e.\ when $\det{(A)}$ vanishes. However, instead of using Eq.  (\ref{C1-C2}) for determining the tensor coefficients $C_1$ and $C_2$, we can assume that these coefficients still exist and express $C_0$ in terms of two-point functions by writing $p_2^2R_1 -p_1p_2R_2 = 0$ and $-p_1p_2R_1 + p_1^2R_2 = 0$ and solving these (equivalent) equations for $C_0$.

The main idea of Ref.\ \cite{Ganesh} is to repeat this procedure for every tensor rank successively. We can write down the expressions determining the tensor coefficients of second rank, i.e.\ $C_{00}$, $C_{11}$, $C_{12}$ and $C_{22}$. The ultraviolet divergent coefficient $C_{00}$ is not directly\footnote{The coefficient $C_{00}$ is indirectly plagued by problems in the limit of a vanishing Gram determinant, as it is composed of the problematic coefficients $C_1$ and $C_2$.} affected by problems of vanishing Gram determinants and found to be

\beq
C_{00} = \frac{m_0^2C_0}{D-2} + \frac{B_0(1,2) + f_1C_1 + f_2C_2}{2(D-2)}.
\eeq
\\

\noindent
In contrast, the remaining tensor coefficients can not be obtained via standard tensor reduction for vanishing Gram determinant. Instead of that, the corresponding equations can be used to determine the tensor equations of rank one, i.e.\ $C_1$ and $C_2$. The result can be written in compact form as 

\beq
\begin{pmatrix} C_1 \\ C_2 \end{pmatrix} = Z_i^{-1}\begin{pmatrix} R_{3,i} \\ R_{4,i}\end{pmatrix}\quad\mathrm{with}\quad i=1,2.\label{C1C2Alternative}
\eeq
\\

\noindent
The abbreviations used here are

\begin{widetext}
\bea
R_{3,i} & = & x_{i1}\left(B_1(1,2) + B_0(1,2) -\frac{2m_0^2}{D-2}C_0 - \frac{1}{D-2}B_0(1,2)\right) + x_{i2}\left(B_1(0,1) + B_1(1,2) + B_0(1,2)\right),\\
R_{4,i} & = & x_{i1}\left(B_1(0,2) - B_1(1,2)\right) + x_{i2}\left(-B_1(1,2) -\frac{2m_0^2}{D-2}C_0 - \frac{1}{D-2}B_0(1,2)\right),\\
Z_i & = &\begin{pmatrix} Y_i + \frac{x_{i1}}{D-2}f_1 & \frac{x_{i1}}{D-2}f_2 \\ \frac{x_{i2}}{D-2}f_1 & Y_i + \frac{x_{i2}}{D-2}f_2 \end{pmatrix}\quad\mathrm{with}\quad Y_i = x_{i1}f_1 + x_{i2}f_2\quad\mathrm{and}\quad \begin{pmatrix} p_2^2 & -p_1p_2\\ -p_1p_2 & p_1^2\end{pmatrix} = \begin{pmatrix} x_{11} & x_{12}\\ x_{21} & x_{22}\end{pmatrix}.
\eea
\end{widetext}

To summarize, the presented method allows to determine the tensor coefficients of rank $n$ by investigating the equations for tensor coefficients of rank $n+1$ in the limit of vanishing Gram determinant. This works in an algorithmic manner. In comparison to the standard tensor reduction method, the expressions are more lengthy. However, note that the algebraic form of Eqs.\ (\ref{C1-C2}) and (\ref{C1C2Alternative}) is the same. One might ask what happens when $\det(Z_i)$ vanishes. This is of interest, as we precisely run into this situation in the course of our direct detection calculations when evaluating e.g.\ the three-point function $C_0(p,p,m_0^2,m_1^2,m_2^2)$.

\begin{figure*}
	\centering
		\includegraphics[width=0.49\textwidth]{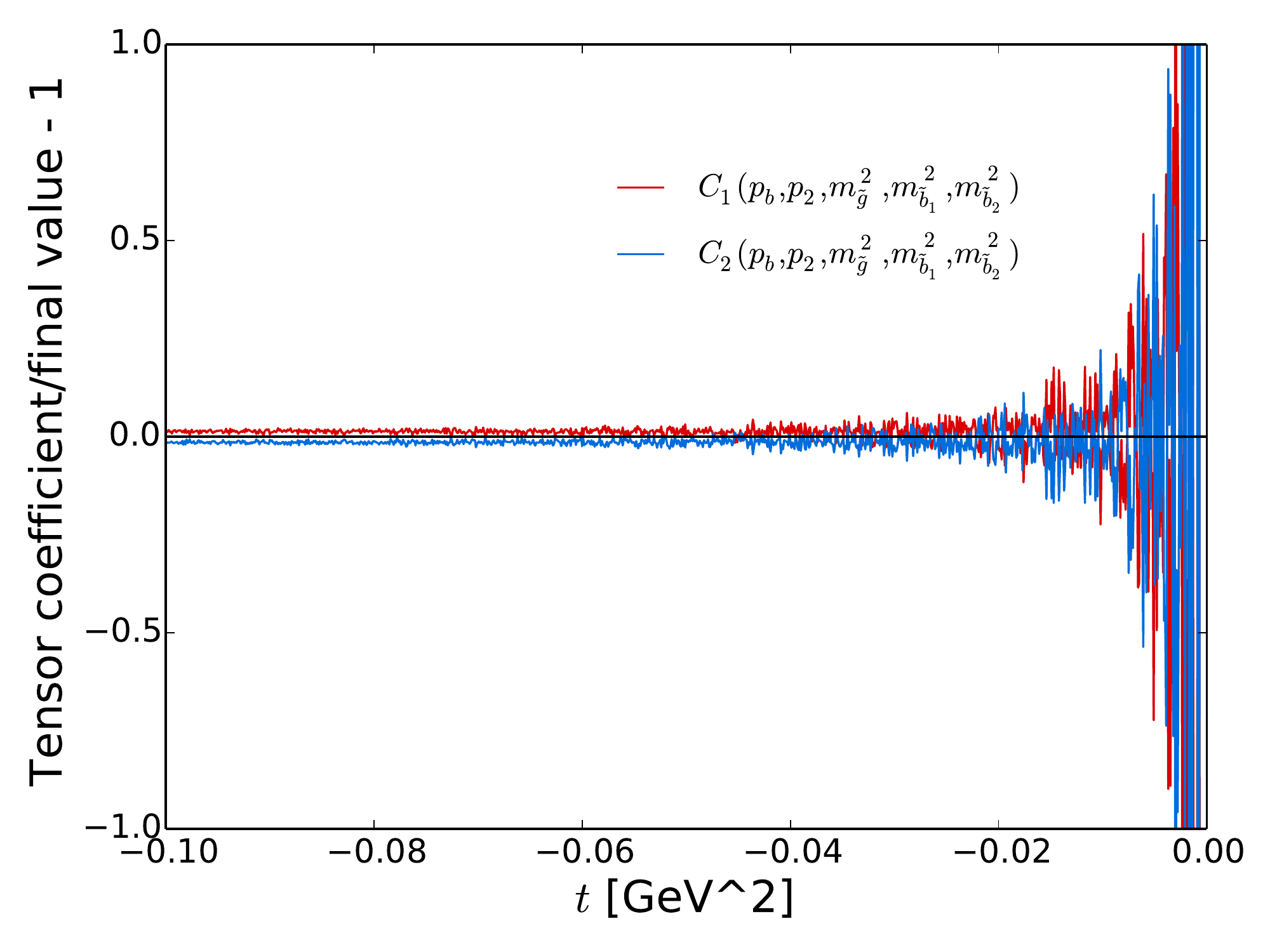}
		\includegraphics[width=0.49\textwidth]{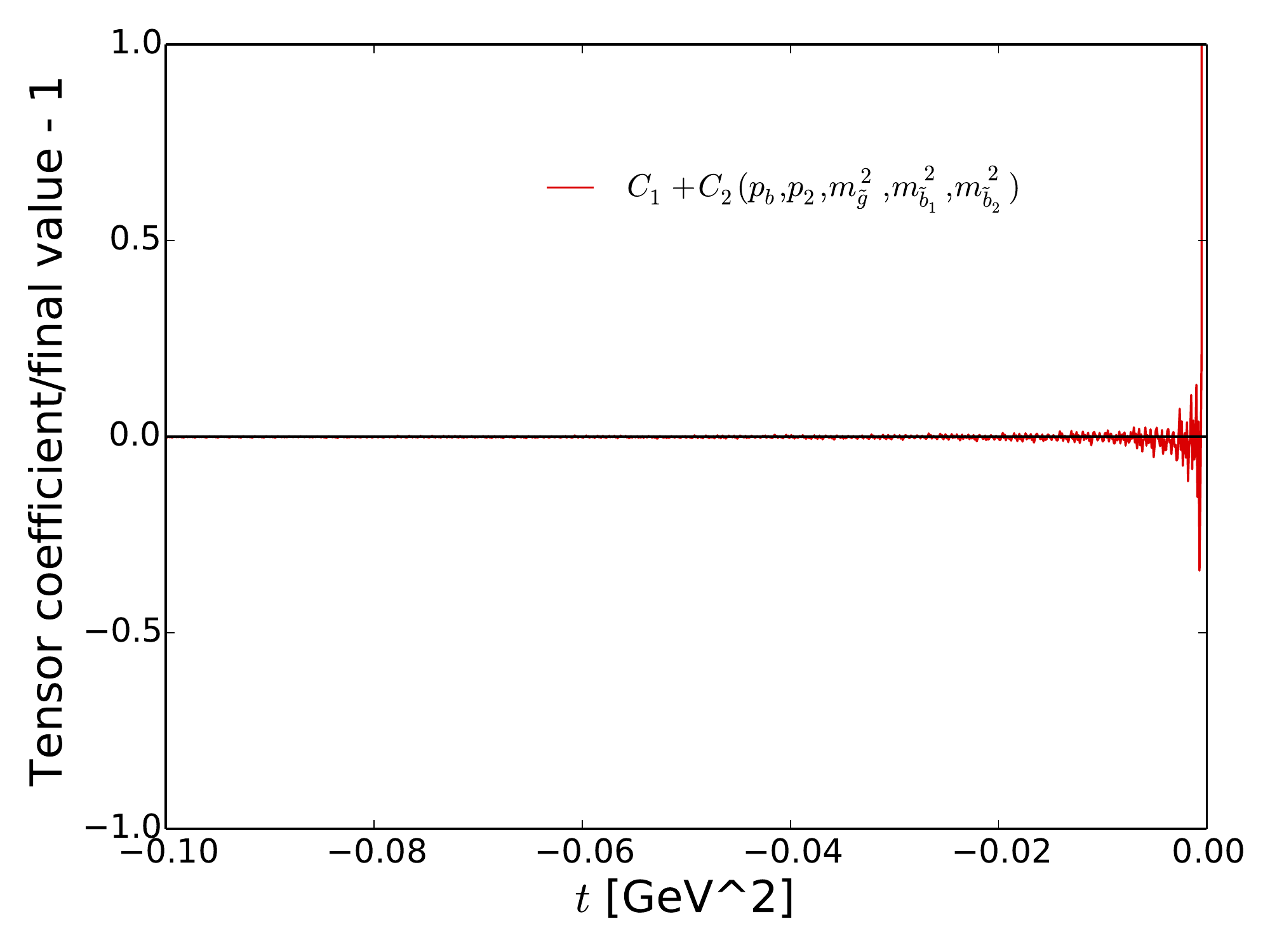}
		\includegraphics[width=0.49\textwidth]{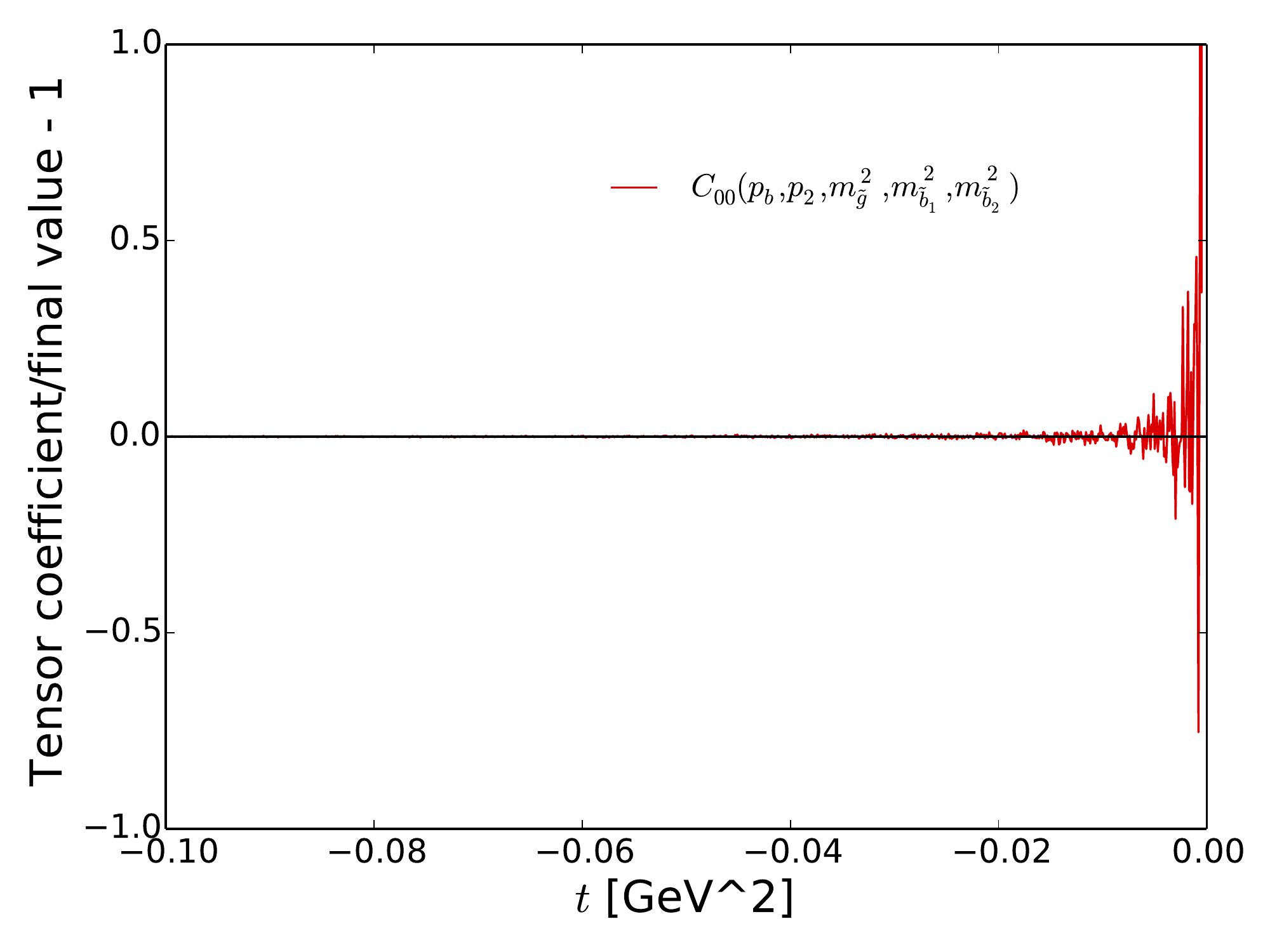}
		\includegraphics[width=0.49\textwidth]{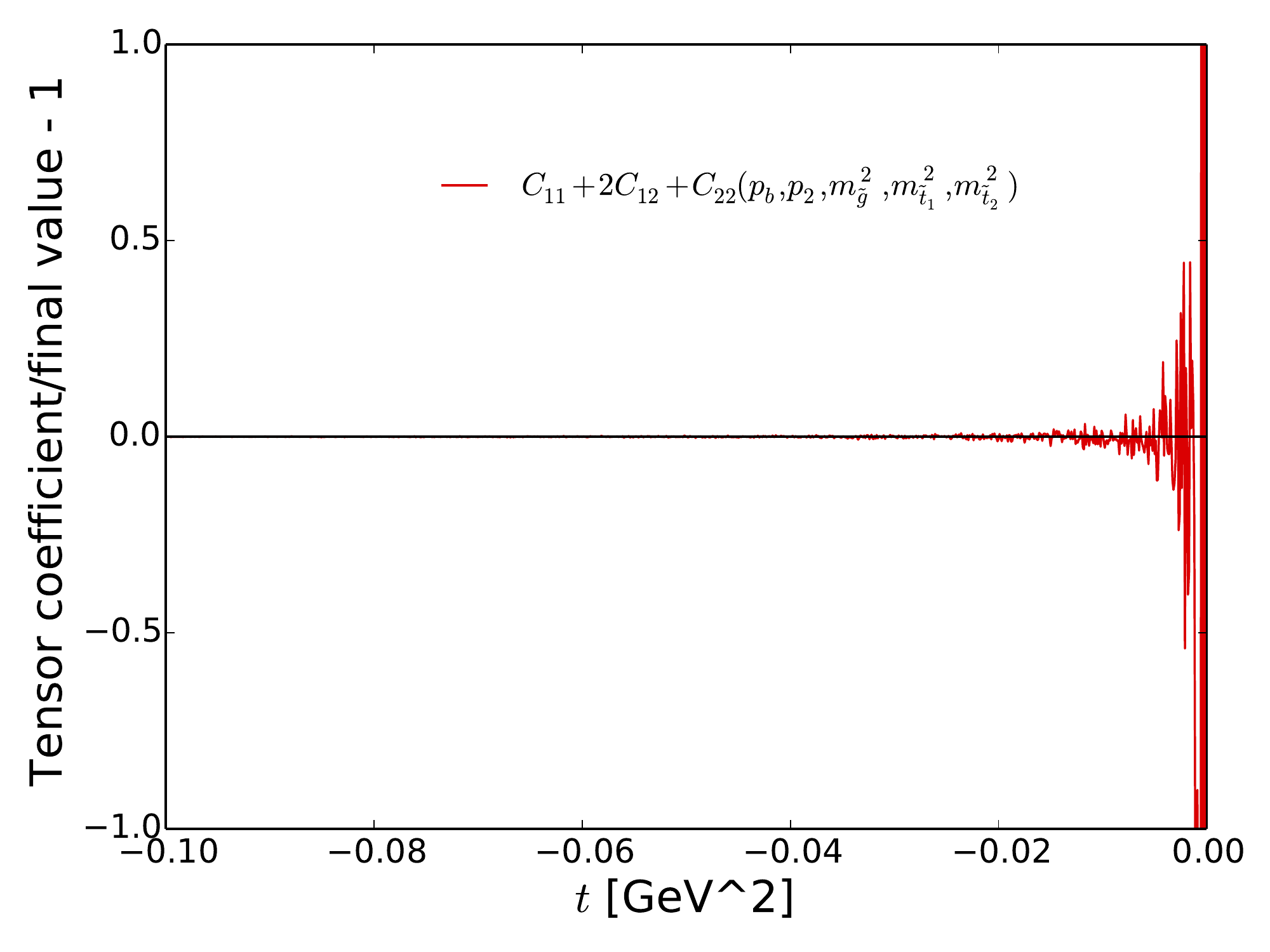}
	\caption{Numerical stability of the three-point tensor coefficients in the limit $v\rightarrow 0$ or equivalently $t\rightarrow 0$.}
	\label{fig:AltReduction}
\end{figure*}

There are basically three ways to proceed. First, note that there are actually two sets of equations hiding behind Eq.\ (\ref{C1C2Alternative}). In some lucky cases it might happen that only one variant fails while the other is still working. The second possibility is to apply l'H\^opital's rule without encountering $\det(Z_i)$ again. This is an improvement in comparison to the standard tensor reduction method, where $\det(A)$ usually reappears when taking the limit. 

We took a closer look at the problematic cases involved in our calculation and found a third way out of this dilemma. We illustrate this by referring to the three-point function $C_0(p,p,m_0^2,m_1^2,m_2^2)$. Remember that $p_1 = p_2 = p$ is a stronger condition than just $p_1^2 = p_2^2 = p^2$ which happens frequently for identical external particles. The crucial observation is that the tensor coefficients $C_1$ and $C_2$ are no longer uniquely defined in this situation; only their sum $C_a = C_1 + C_2$ is. Instead of Eq.\ (\ref{StandardDecomposition}), we get
\beq
C_\mu  =  p_{1,\mu}C_1 + p_{2,\mu}C_2 \rightarrow p_\mu(C_1 + C_2) = p_\mu C_a.
\eeq
It is precisely this combination which remains in all the amplitudes in the limit $p_1\rightarrow p_2$. Hence we replace this sum by $C_a$. This coefficient can be easily obtained in the usual way and reads
\beq
C_a =  \frac{1}{2p^2}\left(B_0(0,2) - B_0(1,2) - f_1C_0\right) \label{Ca}.
\eeq

By taking into account that tensor coefficients may coalesce under certain kinematical circumstances and applying the method of Ref.\ \cite{Ganesh} as illustrated above, we were able to stabilize the tensor reduction method for vanishing Gram determinant for all loop amplitudes occuring in our direct detection analysis. This is particularly true for the four-point functions needed for the box contributions. Although the basic idea remains unchanged, the corresponding expressions become very large and were therefore calculated with the help of \texttt{Mathematica}.

All tensor coefficients obtained in this way have been tested extensively. We have numerically compared them with the corresponding coefficients resulting from the standard tensor reduction method for small, but non-vanishing Gram determinant. Some examples are shown in Fig.\ \ref{fig:AltReduction}.
The upper left plot of Fig.\ \ref{fig:AltReduction} shows the numerical stability of the tensor coefficients $C_1$ and $C_2$ in the limit of equal momenta or equivalently $(p_b - p_2)^2 = t \rightarrow 0$. More precisely, we show the real parts of $C_1$ (in red) and $C_2$ (in blue) obtained by the regular tensor reduction method divided by $C_a/2$ obtained via Eq.\ (\ref{Ca}) and subtracted by one. In this representation, the black null line corresponds directly to $C_a/2$. We observe, as expected, that both $C_1$ and $C_2$ are relatively stable for $t \leq -0.5$ GeV$^2$ and marginally differ from $C_a/2$. The regular tensor reduction is still working here and the small, but finite velocity leads to a small shift relative to the black reference line. However, when we approach the limit $t\rightarrow0$ the regular tensor method fails and both of the coefficients become numerically unstable.

As explained before, the individual tensor coefficients $C_1$ and $C_2$ are no longer uniquely defined in this limit, only their sum is. The real part of this sum divided by $C_a$ and subtracted by one is shown in the upper right plot of Fig.\ \ref{fig:AltReduction}. It is more stable than the individual coefficients, but still becomes noisy at very small relative velocities. For larger (but still small) relative velocities, the agreement between $C_1 + C_2$ and $C_a$ is excellent, which justifies our approach.

We show analogous plots for the tensor coefficient $C_{00}$ and the combination $C_{11} + 2C_{12} + C_{22}$ in the lower part of Fig.\ \ref{fig:AltReduction}. The main features are similar to the ones discussed before. Using the original tensor reduction method, the tensor coefficients become numerically unstable at very small relative velocities. When using the alternative approach described in this section, we obtain a stable result for $v = 0$ which is in perfect agreement with the standard method for small, but non-zero relative velocities. The black reference line in the lower right plot of Fig.~\ref{fig:AltReduction} is defined by
\bea
C_b & = &\frac{1}{3p^2}\big(B_0(1,2) - m_0^2C_0 + 2B_1(0,1)\nonumber\\
&& - 2f_2C_a - \frac{1}{2}\big)
\eea

Although of minor importance for the tensor reduction itself, we list all the masses used in the plots above for completeness. They are $m_b = 2.3$ GeV, $m_t = 148.0$ GeV, $m_{\tilde{g}} = 1170.7$ GeV, $m_{\tilde{b}_1} = 1007.3$ GeV, $m_{\tilde{b}_2} = 1071.9$ GeV, $m_{\tilde{t}_1} = 827.9$ GeV and finally $m_{\tilde{t}_2} = 1042.6$ GeV. Note that we have $p_2^2 = p_b^2 = m_b^2$ in the first three plots, whereas $p_2^2 = p_b^2 = m_t^2$ in the lower right plot. More details on the presented alternative tensor reduction method can be found in Ref.\ \cite{myPhD}.


%% file: bib.tex
\bibliographystyle{apsrev}